\documentclass[twocolumn,showpacs,preprintnumbers,amsmath,amssymb]{revtex4-1}
\usepackage{graphicx}
\usepackage{dcolumn}
\usepackage{bm}
\usepackage{braket}
\usepackage[usenames,dvipsnames]{color}
\usepackage{ulem}

\begin{document}

\title{Electric and magnetic multipoles and bond orders in excitonic insulators}

\author{
Tatsuya Kaneko 
and Yukinori Ohta}
\affiliation{
Department of Physics, Chiba University, Chiba 263-8522, Japan
}

\date{\today}

\begin{abstract}
We study the charge and spin density distributions of excitonic insulator (EI) states in the tight-binding approximation. 
We first discuss the charge and spin densities of the EI states when the valence and conduction bands are composed of orthogonal orbitals in a single atom. 
We show that the anisotropic charge or spin density distribution occurs in a unit cell (or atom) and a higher rank electric or magnetic multipole moment 
becomes finite, indicating that the EI state corresponds to the multipole order.  A full description of the multipole moments for the $s$, $p$, and $d$ orbitals 
is then given in general.  We find that, in contrast to the conventional density-wave states, the modulation of the total charge or net magnetization does not 
appear in this case.  However, when the conduction and valence bands include the component of the same orbital, the modulation of the total charge or net 
magnetization appears, as in the conventional density-wave state.  We also discuss the electron density distribution in the EI state when the valence and 
conduction bands are composed of orbitals located in different atoms.  We show that the excitonic ordering in this case corresponds to the bond order formation.  
Based on the results thus obtained we discuss the EI states of real materials recently reported.  
\end{abstract}

\pacs{71.10.Fd, 71.30.+h, 71.35.Lk, 75.10.-b} 

\maketitle
\section{Introduction}
The formation and condensation of excitonic bound states of electrons and holes in a small band-overlap semimetal or 
a small band-gap semiconductor were predicted theoretically half a century ago~\cite{Mo61,Kn63}.  
The excitonic phase, often referred to as the excitonic insulator (EI), is described by the quantum condensation 
of such excitons triggered by the interband Coulomb interaction 
\cite{Mo61,Kn63,KK65,De65,KM65,KM66,KM66-2,JRK67,Ko67,HR68,HR68-2,Zi67,Zi67-2,Zi68,Zi68-2,KK68,KS70,GK73,VKR75}.  
The excitonic condensation in semimetallic systems can be described in analogy with the BCS theory of superconductors, 
and that in semiconducting systems can be discussed in terms of the Bose-Einstein condensation (BEC) of preformed 
excitons~\cite{CN82,NC82,NS85,BF06}.  
It is also known that, when the valence band top and conduction band bottom are separated by the wave vector $\bm{Q}$, 
the system shows the density wave with modulation $\bm{Q}$~\cite{JRK67,Ko67,HR68,HR68-2}.  
Then, the spin-singlet and spin-triplet EI states are often referred to as excitonic charge-density-wave (CDW) and excitonic 
spin-density-wave (SDW) states, respectively~\cite{Ko67,HR68,HR68-2}.  

Recently, a number of candidate materials for the excitonic phases have been reported, and therefore, the physics of 
excitonic phases has attracted renewed experimental and theoretical attention.  
The candidate materials include the following: 
Tm(Se,Te) was argued to exhibit a pressure-induced excitonic instability, where an anomalous increase in the electrical resistivity 
and thermal diffusivity was reported~\cite{BSW91,BF06}.  
In Ca$_{1-x}$La$_x$B$_6$, the observed weak ferromagnetism was interpreted in terms of the doped spin-triplet EI state~\cite{YHTetal99,ZRA99,BV00,Ba00}.   
The CDW state observed in $1T$-TiSe$_2$ was claimed to be of the excitonic origin~\cite{QHWetal07,CMCetal07,MCCetal09,MSGetal10,WNS10,MBCetal11,MMAetal12,ZFBetal13,MMHetal15,WSY15}.  
Likewise, the structural phase transition observed in a layered chalcogenide Ta$_2$NiSe$_5$ was attributed to the formation 
of a spin-singlet EI state~\cite{WSTetal09,WSTetal12,KTKetal13,SWKetal14,SKO16,YDO16,DYO16}.  
The SDW states of chromium~\cite{Ri70,So78,Fa88,GVA02} and iron-based superconductors 
\cite{MSW08,CEE08,HCW08,VVC09,BT09-1,BT09-2,EC10,KECetal10,VVC10,FS10,KEM11,ZTB11,FCKetal12,SBT14} 
were sometimes argued to be of the excitonic origin as well.  
The condensation of spin-triplet excitons was also predicted to occur in the proximity of the spin state transition, 
of which cobalt oxide materials with perovskite structure are an example~\cite{KA14-2,Ku14,INTetal16,ILTetal16,SK16,TMNetal16}.  

The EI states in strongly correlated electron systems have in particular been addressed in recent years~\cite{Ku15,Ka16}. 
From the theoretical standpoint, the extended Falicov-Kimball model was studied extensively in the context of 
the EI \cite{Ba02,BGBetal04,SC08,Fa08,IPBetal08,Br08,ZFB10,ZIBetal10,PBF10,ZIBetal11,PFB11,SEO11,ZIBetal12,KEFetal13,EKOetal14,FC14}. 
Although this model is the simplest to realize the EI, the spin degrees of freedom are not included in it.  
The EI states with spin degrees of freedom were then discussed in terms of the two-band Hubbard model~\cite{KSO12,KA14-1,KO14,KZFetal15,NWNetal16,KG16}.
It is known that the spin-singlet and spin-triplet EI states in the two-band Hubbard model are exactly degenerate 
when only the interband direct Coulomb interaction is taken into account, and moreover that the spin-triplet EI state is stabilized 
when the interband exchange interactions, such as Hund's rule coupling, are taken into account~\cite{KA14-1,KO14,KZFetal15}.  
The importance of electron-phonon couplings was also pointed out for 1$T$-TiSe$_2$ and Ta$_2$NiSe$_5$~\cite{MBCetal11,ZFBetal13,KTKetal13}; 
the studies of the multiband models with electron-phonon coupling showed that the spin-singlet EI states are actually stabilized by the strong electron-phonon coupling~\cite{PBF13,ZFB14,WSY15,KZFetal15}.  

However, not much is known about the charge density $\rho(\bm{r})$ and spin density $\bm{s}(\bm{r})$ distributions in the EI states 
because they include spatial variations of the Bloch wave functions $\psi_{\bm{k}n}(\bm{r})$ of the systems~\cite{HR68-2,VKR75,BT09-2}.  
This is in particular the case when we consider the EI states in strongly correlated electron systems; 
on the one hand, the tight-binding lattice models were studied much in detail using sophisticated numerical techniques to show the presence of the EI states, 
but on the other hand, their electron density distributions caused by the spatial variations of the Bloch functions were not sufficiently worked out.  
For orbital diagonal orders, such as antiferromagnetism and charge orders in a single-band Hubbard model, 
we need not pay much attention to the Bloch functions because the total charge and magnetization in a unit cell (or atom) 
can be characterized by the square of the same local wave function. 
However, for orbital off-diagonal orders such as the EIs, the deviation in the local charge and spin density distributions 
occurs due to the spontaneous hybridization between different local orbitals.  Here, the anisotropic electron distribution is caused by 
the product of the different local orbitals and therefore becomes highly complicated depending on the spatial position $\bm{r}$ 
in a unit cell (or atom).  
To elucidate the electronic structure of the EIs, we therefore need to evaluate the local charge density $\rho(\bm{r})$ 
and local spin density $\bm{s}(\bm{r})$ from the local wave functions in a unit cell (or atom).  

In this paper, motivated by the above developments in the field, we study the charge and spin density distributions 
in the spin-singlet and spin-triplet EI states, where we fully take into account the spatial shapes of the local (atomic) wave functions 
in the tight-binding approximation.  
We will first discuss the charge and spin density distributions of the EI state when the valence and conduction bands are composed of orthogonal orbitals in a single atom.  
We will show that the anisotropic distribution of the charge or spin density occurs in each unit cell (or in each atom) and 
a higher rank electric or magnetic multipole moment becomes finite, depending on the wave functions of orbitals in the valence and conduction bands. 
The EI states thus correspond to the multipole orders, for which we will give a full description of the multipole moments for the $s$, $p$, and $d$ orbitals in general.  
We will emphasize that, in contrast to the conventional density-wave states, the modulation of the total charge (electric monopole moment) 
or net magnetization (magnetic dipole moment) in each atom does not appear when the orthogonal two orbitals are hybridized via the spin-singlet or 
spin-triplet excitonic ordering.  
However, if the conduction and valence bands include the same orbital component, the density-wave modulation similar to the conventional density-wave states appears.  
We will furthermore discuss the electron density distributions in the EI states when the valence and conduction bands are composed of orbitals in different atoms.  
In this case, the excitonic ordering induces the spontaneous electron bonding between the two orbitals in the different atoms, 
and therefore the EI state corresponds to the bond order formation.  

The rest of this paper is organized as follows:  
In Sec.~II, we will briefly review the theory of the EI state.  
In Sec.~III, we will discuss the electronic density distributions in the  EIs when the valence and conduction bands are composed of orbitals in a single atom.  
We will also discuss the description of the EI state in terms of the electric or magnetic multipole moments.   
In Sec.~IV, we will discuss the electronic density distributions in the EI state when the valence and conduction bands come from orbitals in different atoms.  
In Sec.~V, we will discuss implications of our results 
in the materials aspects, whereby we will draw conclusions.

\section{Excitonic Insulator State}

Let us briefly review the theory of EI here, considering one of the simplest models that describe the EI state, 
which is defined by the Hamiltonian 
\begin{align}
\mathcal{H} = \sum_{\bm{k}}\sum_{\sigma} \varepsilon_{a} (\bm{k})a^{\dag}_{\bm{k}\sigma}a_{\bm{k}\sigma} +  \sum_{\bm{k}}\sum_{\sigma} \varepsilon_{b}(\bm{k}) b^{\dag}_{\bm{k}\sigma}b_{\bm{k}\sigma}  \notag \\
+ \frac{V}{N} \sum_{\bm{k},\bm{k}',\bm{q}}\sum_{\sigma,\sigma'} b^{\dag}_{\bm{k}+\bm{q}\sigma}b_{\bm{k}\sigma}a^{\dag}_{\bm{k}'-\bm{q}\sigma'}a_{\bm{k}'\sigma'},
\label{ei-eq1}
\end{align} 
where $a^{\dag}_{\bm{k}\sigma}$ ($a_{\bm{k}\sigma}$) and $b^{\dag}_{\bm{k}\sigma}$ ($b_{\bm{k}\sigma}$) 
denote the creation (annihilation) operator of an electron with spin $\sigma$ in the valence and conduction bands, respectively, 
and $\varepsilon_a(\bm{k})$ and $\varepsilon_b(\bm{k})$ are their band dispersions.  
$V$ is the interband Coulomb interaction, for which we consider only the on-site Coulomb repulsion 
$V \sum_{i}\sum_{\sigma,\sigma'} n_{i b \sigma} n_{i a \sigma'} = V \sum_{i}\sum_{\sigma,\sigma'} b^{\dag}_{i\sigma}b_{i\sigma}a^{\dag}_{i\sigma'}a_{i\sigma'} $. 
This type of interaction is included in the spinless extended Falicov-Kimball and multiband Hubbard models and is known to drive the system into the EI state.  
Although the multiband Hubbard model includes the intraband Coulomb, Hund's rule coupling, and pair-hopping interactions as well, 
the dominant term inducing the excitonic phase is the interband direct Coulomb interaction $V$~\cite{HR68-2}.  
We therefore consider only the $V$ term for simplicity, thereby describing the EI state.  

The order parameter of the EI state is given by $\langle b^{\dag}_{\bm{k}+\bm{Q}} a_{\bm{k}}\rangle$ 
when the valence band top and conduction band bottom are separated by the wave vector $\bm{Q}$.  
Taking into account the spin degrees of freedom, we can assume either the spin-singlet or spin-triplet electron-hole pairing.  
The order parameters are then defined by
\begin{align}
\Delta_s =& -\frac{V}{2N}\sum_{\bm{k}} \sum_{\sigma} \langle b^{\dag}_{\bm{k}+\bm{Q}\sigma} a_{\bm{k}\sigma}\rangle 
\label{ei-eq2} \\
\bm{\Delta_t} =& -\frac{V}{2N}\sum_{\bm{k}} \sum_{\sigma,\sigma'} \langle b^{\dag}_{\bm{k}+\bm{Q}\sigma} \bm{\sigma}_{\sigma\sigma'}a_{\bm{k}\sigma'}\rangle ,
\label{ei-eq3}
\end{align} 
for the spin-singlet and spin-triplet EI states, respectively, 
where $\bm{\sigma}=(\sigma^x,\sigma^y,\sigma^z)$ is the vector of Pauli matrices~\cite{HR68-2}.  
$\bm{\Delta_t} = (\Delta^x_t, \Delta^y_t, \Delta^z_t)$ is the vector order parameter, reflecting the spin direction. 

If we assume a direct-gap system ($\bm{Q}=0$) and apply the mean-field approximation for simplicity, we obtain the mean-field Hamiltonian as 
\begin{align}
\mathcal{H}(\bm{k}) = \left(
   \begin{array}{cc}
     \varepsilon_a(\bm{k})I & \Delta_s  I + \bm{\Delta}_t \cdot\bm{\sigma}\\
      \Delta_s^{*}I + \bm{\Delta}_{t}^*\cdot\bm{\sigma} & \varepsilon_b(\bm{k})I 
   \end{array}
  \right), 
\label{ei-eq4} 
\end{align}
where we use the basis $(a^{\dag}_{\bm{k}\uparrow}\; a^{\dag}_{\bm{k}\downarrow}\; b^{\dag}_{\bm{k}\uparrow}\; b^{\dag}_{\bm{k}\downarrow})$ 
and $I$ is the unit matrix~\cite{VKR75,KA14-1}.  The order parameters are calculated self-consistently to obtain solutions with $\Delta_s\ne 0$ or $\bm{\Delta}_t \ne0$.  
It has been confirmed that the EI states actually appear in the extended Falicov-Kimball and two-band Hubbard models, 
where not only the mean-field approximation but also more accurate numerical methods were used~\cite{BGBetal04,SEO11,KEFetal13,EKOetal14,KSO12,KA14-1}.  

It is known that the spin-singlet and spin-triplet EI states are energetically degenerate in the model of Eq.~(\ref{ei-eq1}), 
where only the interband direct Coulomb interaction is considered~\cite{HR68-2,KZFetal15}.  
This degeneracy is lifted if we take into account other interactions: 
the spin-triplet EI is stabilized by the interband exchange interactions, such as Hund's rule coupling, and 
the spin-singlet EI state is stabilized by a strong electron-phonon coupling~\cite{HR68-2,KZFetal15}.  
The order parameters are in general complex and the energy of the system is independent of the choice of the phase of the order parameters. 
It is known however that the phase is fixed by the pair-hopping interaction and/or electron-phonon coupling, 
such that the order parameters are taken to be real~\cite{HR68-2,ZFB14,KZFetal15,NWNetal16}. 
We thus assume the real order parameters in this paper.  

When the valence and conduction bands are hybridized spontaneously due to the EI state formation, 
where the order parameters $\Delta_s $ or $\bm{\Delta_t}$ becomes finite, the change in the local charge 
or spin density distributions is given by the band off-diagonal expectation values 
$\langle b^{\dag}_{\bm{k}+\bm{Q}\sigma} a_{\bm{k}\sigma}\rangle \ne 0$ or 
$\langle b^{\dag}_{\bm{k}+\bm{Q}\sigma} \bm{\sigma}_{\sigma\sigma'}a_{\bm{k}\sigma'}\rangle \ne 0$.  
The physical meaning of the EI state was given by Halperin and Rice~\cite{HR68-2}, which we essentially follow 
for the description of the charge and spin density distributions.  
In the two-band model, the field operator of annihilating an electron is given by 
\begin{align}
\Psi _\sigma (\bm{r}) = \sum_{\bm{k}} \left[ \psi_{\bm{k}a}(\bm{r}) a_{\bm{k}\sigma} + \psi_{\bm{k}b}(\bm{r}) b_{\bm{k}\sigma} \right],  
\label{ei-eq5}
\end{align} 
where $\psi_{\bm{k}a}(\bm{r})$ and $\psi_{\bm{k}b}(\bm{r})$ are the Bloch functions of 
the valence and conduction bands, respectively~\cite{HR68-2,VKR75,BT09-2,Ba00}.  
Using this operator, the local charge and spin densities are given by 
\begin{align}
\rho(\bm{r}) &= \sum_{\sigma}\langle \Psi^{\dag}_{\sigma}(\bm{r}) \Psi_{\sigma}(\bm{r}) \rangle,   
\label{ei-eq6} \\
\bm{s} (\bm{r}) &= \frac{1}{2}\sum_{\sigma,\sigma'}\langle \Psi^{\dag}_{\sigma}(\bm{r})  \bm{\sigma}_{\sigma\sigma'} \Psi_{\sigma'}(\bm{r}) \rangle, 
\label{ei-eq7}
\end{align} 
respectively~\cite{HR68-2,VKR75,BT09-2,Ba00}. 
When the spin-singlet EI state is realized, the local charge density becomes 
\begin{align}
\rho(\bm{r}) =  \sum_{\bm{k}}& \sum_{\sigma}   \biggl[  |\psi_{\bm{k}a}(\bm{r})|^2 \langle a^{\dag}_{\bm{k}\sigma} a_{\bm{k}\sigma} \rangle  +  |\psi_{\bm{k}b}(\bm{r})|^2 \langle b^{\dag}_{\bm{k}\sigma} b_{\bm{k}\sigma} \rangle   \notag \\
 +  & \left\{ \psi^{*}_{\bm{k}+\bm{Q}b}(\bm{r})  \psi_{\bm{k}a}(\bm{r})  \langle b^{\dag}_{\bm{k}+\bm{Q}\sigma} a_{\bm{k}\sigma} \rangle  +  \mathrm{H.c.} \right\} \biggr] .
\label{ei-eq8}
\end{align} 
Owing to $\langle b^{\dag}_{\bm{k}+\bm{Q}\sigma} a_{\bm{k}\sigma}\rangle \ne 0$ in the EI state, the change in the charge density distribution is given 
by the third term of Eq.~(\ref{ei-eq8}), so that the density wave with modulation vector $\bm{Q}$ appears in the charge density distribution.  
Therefore, the spin-single EI is often referred to as an excitonic CDW~\cite{Ko67,HR68,HR68-2,QHWetal07,CMCetal07}. 
In the same way, the local spin density of the spin-triplet EI is given by 
\begin{align}
\bm{s}(\bm{r}) =  \frac{1}{2} \sum_{\bm{k}} \sum_{\sigma,\sigma'}   \psi^{*}_{\bm{k}+\bm{Q}b}(\bm{r})  \psi_{\bm{k}a}(\bm{r})  \langle b^{\dag}_{\bm{k}+\bm{Q}\sigma} \bm{\sigma}_{\sigma \sigma'} a_{\bm{k}\sigma '} \rangle  +  \mathrm{H.c.} 
\label{ei-eq9}
\end{align} 
Owing to $\langle b^{\dag}_{\bm{k}+\bm{Q}\sigma} \bm{\sigma}_{\sigma \sigma'} a_{\bm{k}\sigma '} \rangle \ne 0$ in the EI state, 
the local spin polarization appears, so that the spin density distribution shows the density wave with modulation vector $\bm{Q}$.  
Therefore, the spin-triplet EI is often referred to as an excitonic SDW~\cite{HR68,HR68-2,BT09-1,BT09-2,KEM11,ZTB11}.   

As seen in Eqs.~(\ref{ei-eq8}) and (\ref{ei-eq9}), the charge and spin densities of the EI include 
the Bloch function $\psi_{\bm{k}n}(\bm{r})$, for which the description is not obvious~\footnote{In Ref.~[{\onlinecite{KO14}}], we assumed the constant Bloch wave functions~\cite{BT09-2,ZTB11,ZFBetal13} for both Eq.~(18) and Fig.~3, whereby we showed the charge and spin density oscillations observed in the density of state (DOS). However, this estimation was not suitable to describe the character of the EI states in detail. In the exact estimation of the local DOS {$N(\bm{r},\omega)$}, we need to use the exact Bloch functions {$\psi_{\bm{k}n}(\bm{r},\omega)$} corresponding to real materials~\cite{VKR75}. Thus, we note that the calculated DOSs in Ref.~[{\onlinecite{KO14}}] do not directly reflect the total charge or net magnetization in each unit cell.}. 
Unambiguous description of the EI states may rely on the wave functions given in real space, 
for which we may assume either a nearly-free-electron approximation or a tight-binding approximation depending 
on the situations of materials concerned.  
Because candidate materials recently suggested to be in the EI state are among transition-metal compounds, 
their electronic structure may be better described by the tight-binding picture rather than by the free-electron--like picture.  
Theoretical studies of the EI in such strongly correlated electron systems also rely on the lattice models, 
such as extended Falicov-Kimball and multiband Hubbard models, rather than the gas models.  
In this paper, we therefore express the Bloch functions in the tight-binding approximation, 
or as a linear combination of the atomic orbitals, and evaluate the charge and spin density 
distributions of the EIs, where the shapes of the atomic orbitals in real space become important.  
In what follows, we will discuss two cases separately: (i) the case where the valence and conduction bands come 
from the orbitals in {\it a single atom} and (ii) the case where they come from {\it different atoms}.

\section{Multiorbitals in a Single Atom}
\subsection{Charge and spin densities} \label{eics-sa}
First, let us consider the case where the valence and conduction bands are composed of orthogonal orbitals in a single atom.  
In the tight-binding approximation, the Bloch function for the $\alpha$ orbital is given as 
\begin{align}
\psi_{\bm{k}\alpha}(\bm{r})
 = \frac{1}{\sqrt{N}}\sum_{i} e^{i\bm{k}\cdot \bm{R}_i} \phi_{\alpha} (\bm{r}-\bm{R}_i), 
\label{eics-sa-eq1}
\end{align} 
where $\phi_{\alpha} (\bm{r})$ is the atomic wave function of the $\alpha$ orbital and $\bm{R}_i$ is the lattice vector. 
Using this $\psi_{\bm{k}\alpha}(\bm{r})$ and applying the Fourier transformation, the field operator is given in real space as 
\begin{align}
\Psi_{\sigma}(\bm{r})
 = \sum_{i}\sum_{\alpha} \phi_{\alpha} (\bm{r}-\bm{R}_i)c_{i\alpha\sigma}, 
\label{eics-sa-eq2}
\end{align} 
where  $c_{i\alpha\sigma}$ ($c^{\dag}_{i\alpha\sigma}$) is the annihilation (creation) operator of an electron at site $i$ and spin $\sigma$ $(=\uparrow,\downarrow)$ in the $\alpha$ orbital~\cite{Ba00}. 
The charge and spin densities are given by Eqs.~(\ref{ei-eq6}) and (\ref{ei-eq7}) using this field operator.  

Let us assume a two-orbital model for simplicity.  Then, using the orbitals of the valence ($a$) and conduction ($b$) bands, 
the field operator in the $i$-th unit cell (or atom) is given as~\cite{Ba00,KA14-1,Ku14}
\begin{align}
\Psi_{i\sigma}(\bm{r}) =  \phi_{ia} (\bm{r})c_{ia\sigma} + \phi_{ib} (\bm{r})c_{ib\sigma}, 
\label{eics-sa-eq3}
\end{align}
where we write $\phi_{i\alpha} (\bm{r})= \phi_{\alpha} (\bm{r}-\bm{R}_i)$.  
We assume $\phi_{i\alpha} (\bm{r})$ to be real, neglecting the spin-orbit coupling.  
We evaluate the charge and spin densities in a unit cell (or atom), neglecting the electronic densities coming from the neighboring cells (or atoms).  
The essential features of the electronic structure may be grasped if the atomic orbitals are well localized, so that the tight-binding approximation is justified.  

Using the field operator in Eq.~(\ref{eics-sa-eq3}) and assuming the spin-singlet EI state, we write the local charge density in the $i$-th unit cell (or atom) as~\cite{Ba00} 
\begin{align}
\rho_i(\bm{r})
& = \sum_{\sigma}\langle \Psi^{\dag}_{i\sigma}(\bm{r}) \Psi_{i\sigma}(\bm{r}) \rangle \notag \\
& = \sum_{\sigma}\sum_{\alpha,\beta} \phi_{i\alpha} (\bm{r}) \phi_{i\beta} (\bm{r}) \langle c^{\dag}_{i\alpha\sigma}c_{i\beta\sigma}\rangle, 
\label{eics-sa-eq4}
\end{align}
where we note that the charge density in the entire space is given approximately as $\rho(\bm{r})\sim \sum_{i} \rho_i(\bm{r})$. 
Defining the orbital diagonal and off-diagonal terms as 
\begin{align}
n_{i\alpha} & = \sum_{\sigma} \langle c^{\dag}_{i\alpha\sigma}c_{i\alpha\sigma}\rangle,  
\label{eics-sa-eq5} \\
\Phi_{is} & = \sum_{\sigma}\langle c^{\dag}_{ib\sigma} c_{ia\sigma} \rangle, 
\label{eics-sa-eq6}
\end{align}
respectively, we write the local charge density in Eq.~(\ref{eics-sa-eq4}) as~\cite{Ba00} 
\begin{align}
\rho_i(\bm{r})  =   
  \phi^{2}_{ia} (\bm{r}) n_{ia} 
+ \phi^{2}_{ib} (\bm{r}) n_{ib} 
+ 2 \phi_{ia} (\bm{r}) \phi_{ib} (\bm{r})\Phi_{is} . 
\label{eics-sa-eq7}
\end{align}
When the EI state has the modulation vector $\bm{Q}$, we have 
\begin{align}
\Phi_{is} = \sum_{\sigma}\langle c^{\dag}_{ib\sigma} c_{ia\sigma} \rangle = \Phi_s \cos \bm{Q}\cdot \bm{R}_i .  
\label{eics-sa-eq8}
\end{align}
Then, from the third term of Eq.~(\ref{eics-sa-eq7}), the deviation in the charge density caused by the excitonic ordering is given by 
\begin{align}
\delta \rho_i(\bm{r})  =   2 \phi_{ia} (\bm{r}) \phi_{ib} (\bm{r})
\Phi_s \cos \bm{Q} \cdot \bm{R}_i .
\label{eics-sa-eq9}
\end{align}
We thus find from Eq.~(\ref{eics-sa-eq9}) that the charge density has the spatial modulation of $\bm{Q}$ in the spin-singlet EI state.  

Assuming the spin-triplet EI state, we write the local spin density in the $i$-th unit cell (or atom) as~\cite{Ba00,KA14-1,Ku14} 
\begin{align}
\bm{s}_i(\bm{r}) 
& =  \frac{1}{2}\sum_{\sigma,\sigma '} \langle \Psi^{\dag}_{i\sigma}(\bm{r})\bm{\sigma}_{\sigma\sigma'} \Psi_{i\sigma '}(\bm{r}) \rangle \notag  \\
& = \frac{1}{2}\sum_{\sigma,\sigma'}\sum_{\alpha,\beta} \phi_{i\alpha} (\bm{r}) \phi_{i\beta} (\bm{r}) \langle c^{\dag}_{i\alpha\sigma}\bm{\sigma}_{\sigma \sigma'}c_{i\beta\sigma'}\rangle, 
\label{eics-sa-eq10}
\end{align} 
where we note that the spin density in the entire space is given approximately as $\bm{s}(\bm{r})\sim \sum_{i} \bm{s}_i(\bm{r})$. 
Defining the orbital diagonal and off-diagonal terms as 
\begin{align}
\bm{m}_{i\alpha} &= \frac{1}{2} \sum_{\sigma,\sigma'} \langle c^{\dag}_{i\alpha\sigma}\bm{\sigma}_{\sigma\sigma'} c_{i\alpha\sigma'}\rangle, 
\label{eics-sa-eq11} \\
\bm{\Phi}_{it} &= \frac{1}{2} \sum_{\sigma,\sigma'} \langle c^{\dag}_{ib\sigma} \bm{\sigma}_{\sigma\sigma'} c_{ia\sigma'} \rangle , 
\label{eics-sa-eq12}
\end{align}
respectively, we write the local spin density in Eq.~(\ref{eics-sa-eq10}) as~\cite{Ba00,KA14-1,Ku14} 
\begin{align}
 \bm{s}_i(\bm{r})  =   
  \phi^{2}_{ia} (\bm{r}) \bm{m}_{ia} 
+ \phi^{2}_{ib} (\bm{r}) \bm{m}_{ib}  
+ 2 \phi_{ia} (\bm{r}) \phi_{ib} (\bm{r})\bm{\Phi}_{it}. 
\label{eics-sa-eq13}
\end{align}
When the EI state has the modulation vector $\bm{Q}$, we have 
\begin{align}
\bm{\Phi}_{it} = \frac{1}{2}\sum_{\sigma,\sigma'} \langle c^{\dag}_{ib\sigma} \bm{\sigma}_{\sigma\sigma'} c_{ia\sigma'} \rangle = \bm{\Phi}_t \cos \bm{Q} \cdot \bm{R}_i . 
\label{eics-sa-eq14}
\end{align}
Then, from the third term of Eq.~(\ref{eics-sa-eq13}), the deviation in the spin density caused by the excitonic ordering is given by 
\begin{align}
\delta \bm{s}_i(\bm{r}) =2 \phi_{ia} (\bm{r}) \phi_{ib} (\bm{r}) \bm{\Phi}_t \cos \bm{Q} \cdot \bm{R}_i .
\label{eics-sa-eq15}
\end{align}
We thus find from Eq.~(\ref{eics-sa-eq15}) that the spin density has the spatial modulation of $\bm{Q}$ in the spin-triplet EI state.  

An important factor arising in the EI state formation is then 
\begin{align}
F(\bm{r}) = \phi_{ia} (\bm{r}) \phi_{ib} (\bm{r}), 
\label{eics-sa-eq16}
\end{align}
which is a product of the wave functions of the $a$ and $b$ orbitals and has either positive or negative part depending on $\bm{r}$.  
The charge and spin densities are therefore spatially varying due to $F(\bm{r})$, showing a variety of distributions in the unit cell (or atom).  
When the parities of the wave functions $\phi_{ia}(\bm{r})$ and $\phi_{ib}(\bm{r})$ are the same, the parity of $F(\bm{r})$ becomes even.  
However, if the wave functions $\phi_{ia}(\bm{r})$ and $\phi_{ib}(\bm{r})$ have different parities, their product $F(\bm{r})$ has an odd parity, 
breaking the space inversion symmetry in the unit cell (or atom).  
Electronic ferroelectricity, which is derived from the broken inversion symmetry of $F(\bm{r})$, has been suggested to occur in the extended Falicov-Kimball model~\cite{POS96-2,Ba02,ZFB10}.  

\begin{figure}[t]
\begin{center}
\includegraphics[width=\columnwidth]{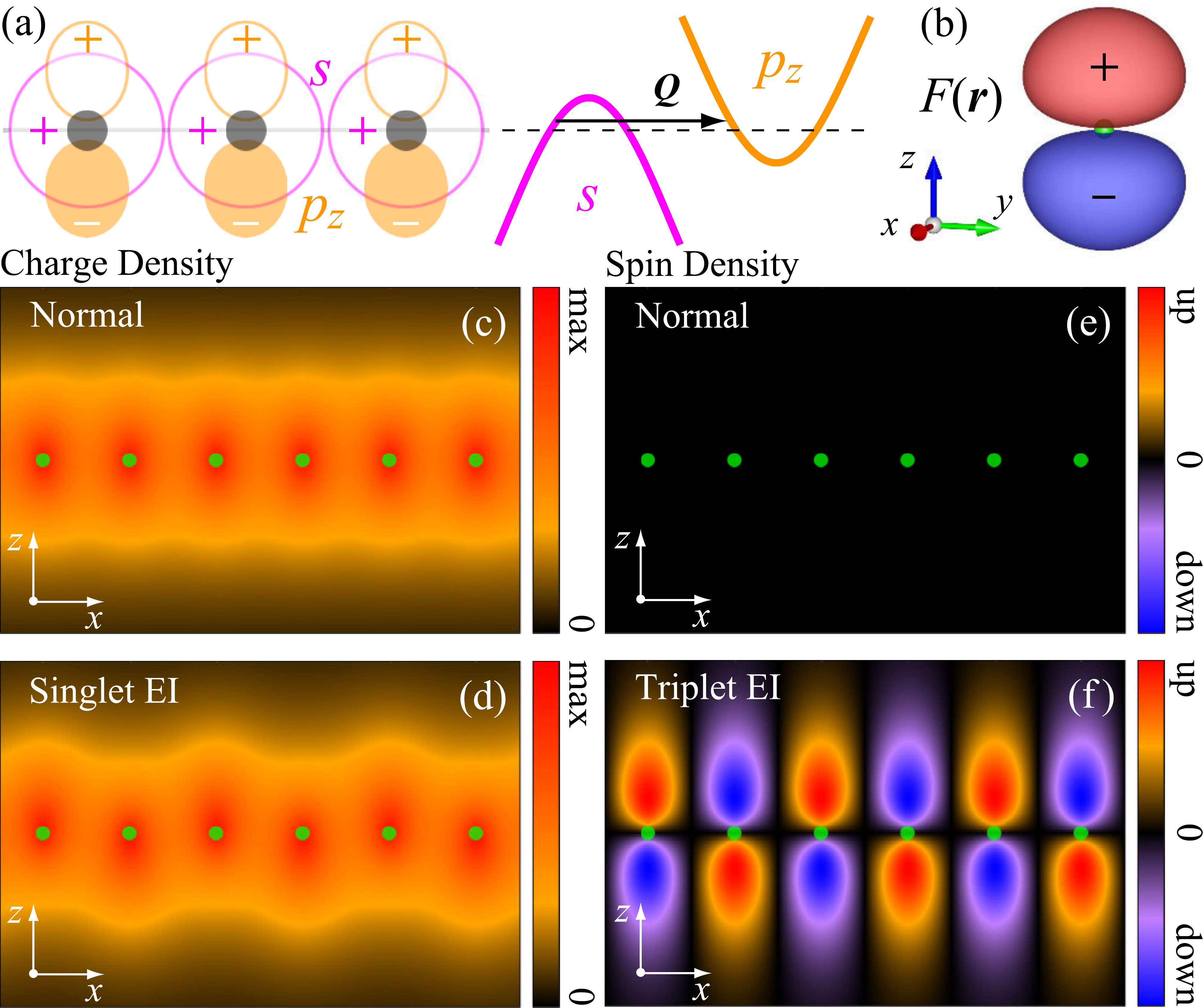}
\caption{(Color online) 
(a) Schematic representations of the valence $s$ and conduction $p_z$ orbitals in the two-dimensional square lattice (side view) and their band dispersions.  
(b) Isosurface of $F(\bm{r})=\phi_{s}(\bm{r})\phi_{p_z}(\bm{r})$.  Positive and negative parts of $F(\bm{r})$ are indicated by red $(+)$ and blue $(-)$, respectively.  
Illustrated in the lower panels are the side views of the two-dimensional plane: 
(c) the charge density in the normal state and 
(d) that in the spin-singlet EI state with $\bm{Q}=(\pi, \pi)$, and 
(e) the spin density in the normal state and 
(f) that in the spin-triplet EI state with $\bm{Q}=(\pi, \pi)$.  
Note that the radial wave function of the 1$s$ orbital is slightly broadened to exaggerate the character of $F(\bm{r})$ 
to illustrate the charge and spin density distributions although the exact spherical harmonics is assumed for the angular dependencies 
of the $s$ and $p_z$ orbitals.  Thus, (c)-(f) are not exact but schematic illustrations.  
}\label{eics-si-fig1}
\end{center}
\end{figure}

Here, let us consider a simple example, where the valence band $a$ and conduction band $b$ are composed of the $s$ and $p_z$ orbitals, respectively, 
which are located in the two-dimensional square lattice [see Fig.~\ref{eics-si-fig1}(a)].  The valence band top and the conduction band bottom 
are separated again by a vector $\bm{Q}=(\pi,\pi)$.  Then, in the two-band Hubbard model with this noninteracting band structure, 
the orbital off-diagonal (or excitonic) orders are realized as below, where the orbital diagonal terms with $n_{is}=n_{s}$, $n_{ip_z} = n_{p_z}$, and $\bm{m}_{is}=\bm{m}_{ip_z}=0$ are found~\cite{KSO12}.
In the case of the spin-singlet EI state, the local charge density is given by
\begin{align}
\rho_i(\bm{r}) & =   
  \phi^{2}_{is} (\bm{r}) n_{s} 
+ \phi^{2}_{ip_z} (\bm{r}) n_{p_z}  \notag \\
&+ 2 \phi_{is}(\bm{r})\phi_{ip_z}(\bm{r})\Phi_s \cos \bm{Q} \cdot \bm{R}_i ,
\label{eics-sa-eq17}
\end{align}
and the local spin density is given by $\bm{s}_i(\bm{r})=0$. 
In the spin-triplet EI state, on the other hand, the local spin density is given by 
\begin{align}
\bm{s}_i(\bm{r}) & =   
  2 \phi_{is}(\bm{r})\phi_{ip_z}(\bm{r})
\bm{\Phi}_t \cos \bm{Q} \cdot \bm{R}_i , 
\label{eics-sa-eq18}
\end{align}
and the local charge density is given by $\rho_i(\bm{r}) = \phi^{2}_{is} (\bm{r}) n_{s} + \phi^{2}_{ip_z} (\bm{r}) n_{p_z}$.

The product of the wave functions of the $s$ and $p_z$ orbitals, $F(\bm{r})=\phi_{is}(\bm{r})\phi_{ip_z}(\bm{r})$, 
is illustrated in Fig.~\ref{eics-si-fig1}(b), where we find that its parity is odd, breaking the spatial inversion symmetry, 
because the wave functions $\phi_{is}(\bm{r})$ and $\phi_{ip_z}(\bm{r})$ are even and odd, respectively.  
Using this function $F(\bm{r})$, we estimate the charge and spin densities in the normal state ($\Phi_s=0$, $\bm{\Phi}_t=0$), 
the charge density in the spin-singlet EI state ($\Phi_s\ne0$), and the spin density in the spin-triplet EI state ($\bm{\Phi}_t\ne0$), 
which are illustrated in Figs.~\ref{eics-si-fig1}(c)--\ref{eics-si-fig1}(f).  
We find that the charge density in the normal state is uniform [see Fig.~\ref{eics-si-fig1}(c)], 
but when the spin-singlet EI state occurs ($\Phi_s\ne0$), it deviates towards the $+z$ direction in the $i$-th site 
and towards the $-z$ direction in the neighboring sites due to $F(\bm{r})$, and thus the charge density in the EI state 
has a period twice as long as the original lattice period [see Fig.~\ref{eics-si-fig1}(d)].
Likewise, we find that there is no spin polarization in the normal state [see Fig.~\ref{eics-si-fig1}(e)], 
but when the spin-triplet EI state occurs ($\bm{\Phi}_t\ne0$), the spin polarization corresponding to the 
spatial variation of $F(\bm{r})$ appears in each unit cell (or atom).  This polarization is inverted 
alternately over the unit cells (or atoms), leading to the spin density with a period twice as long as 
the original lattice period [see Fig.~\ref{eics-si-fig1}(f)].  
We thus confirm that the charge and spin densities have the density waves corresponding to $\bm{Q}$ in the spin-singlet and spin-triplet EI states, respectively.  
However, we have to emphasize here that, although the charge and spin densities are thus modulated, 
the total charge in the unit cell (or atom) does not change [see Fig.~\ref{eics-si-fig1}(d)] 
and the net magnetization in the unit cell (or atom) is zero [see Fig.~\ref{eics-si-fig1}(f)], 
which are quite unlike the situations in the conventional CDW and SDW states.  
We note that, in the excitonic phases formed from orbitals in a single atom, the interorbital exchange interactions 
such as the Hund's rule coupling work to arrange the spins ferromagnetically in each unit cell (or atom), and therefore the spin-triplet EI state 
shown in Fig.~\ref{eics-si-fig1}(f) has lower energy than the spin-singlet EI shown in Fig.~\ref{eics-si-fig1}(d) if there is 
no strong electron-phonon coupling~\cite{HR68-2,KZFetal15}.  

\begin{figure}[t]
\begin{center}
\includegraphics[width=0.9\columnwidth]{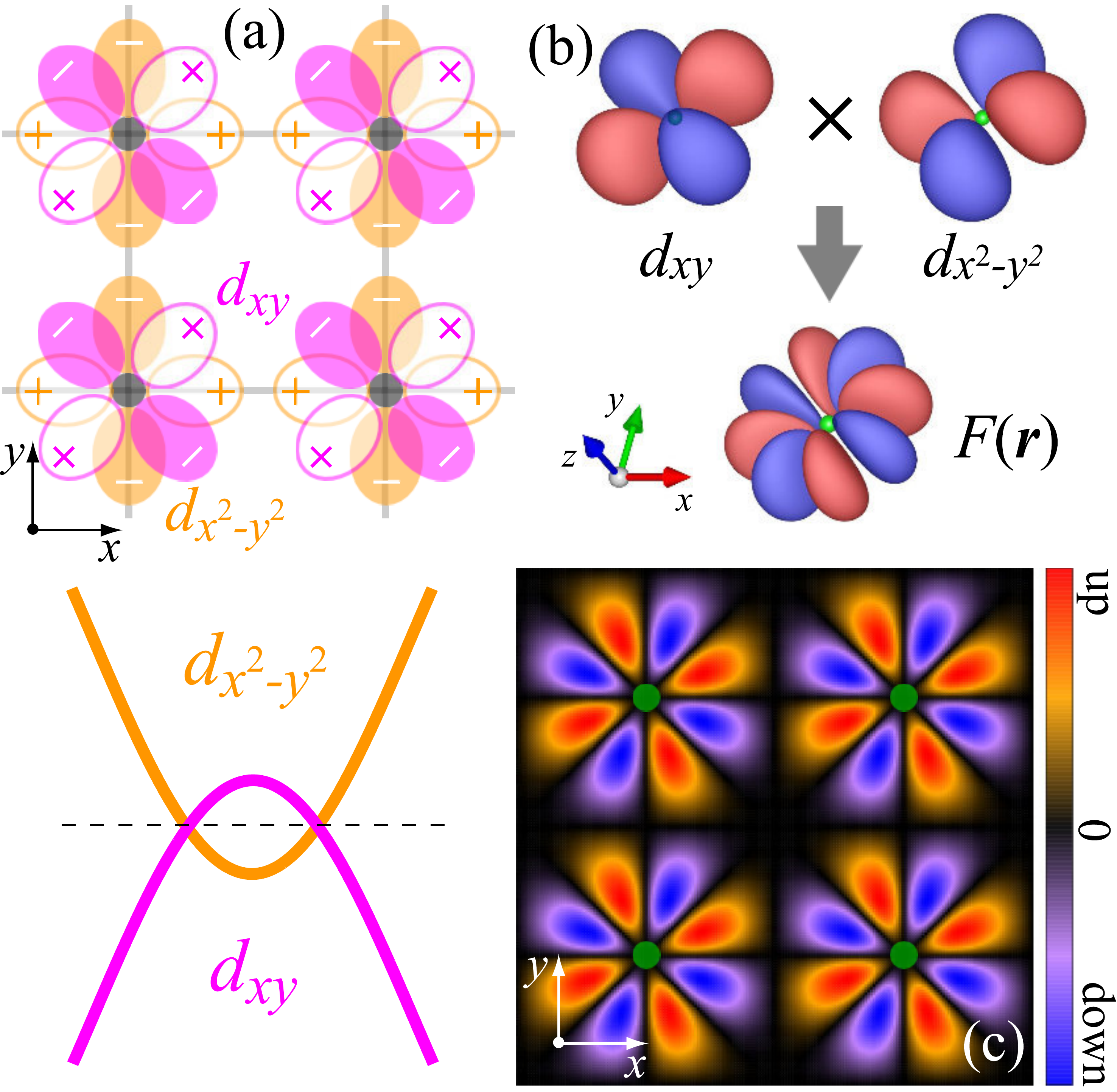}
\caption{(Color online) 
(a) Schematic representations of the valence $d_{xy}$ and conduction $d_{x^2-y^2}$ orbitals in the two-dimensional square lattice (top view) and their band dispersions.  
(b) Isosurfaces of the $d_{xy}$ and $d_{x^2-y^2}$ wave functions and their product $F(\bm{r})$.  
Positive and negative parts of these functions are plotted in different colors.  
(c) Spin density distribution in the corresponding spin-triplet EI state with $\bm{Q}=0$, 
where the up-spin and down-spin distributions are indicated by red and blue, respectively.  
}\label{eics-si-fig2}
\end{center}
\end{figure}

When we consider the strongly correlated electron system, where the wave functions $\phi_{i\alpha}(\bm{r})$ are typically the $d$ or $f$ orbitals, 
we may find a very complicated spatial dependence of $F(\bm{r})$.  In Fig.~\ref{eics-si-fig2}, we show an example where the spin-triplet excitonic ordering 
occurs from the $d_{xy}$ and $d_{x^2-y^2}$ orbitals~\cite{NWNetal16}.  The product of the wave functions $F(\bm{r})$ has eight nodes [see Fig.~\ref{eics-si-fig2}(b)], 
resulting in a complicated spin density distribution in each unit cell (or atom).  
In this example, the hopping integral between the $d_{xy}$ orbitals has the opposite sign to that between the $d_{x^2-y^2}$ orbitals, 
i.e., $t_{xy}t_{x^2-y^2}<0$~\footnote{In the example shown in Fig.~\ref{eics-si-fig2}(a), the hopping integrals between the nearest-neighbor sites are given by $t_{xy}=t(dd\pi)<0$ and $t_{x^2-y^2}=3t(dd\sigma)/4 + t(dd\delta)/4 >0$, where the Slater-Koster parameters \cite{SK54} $t(dd\pi)$ ($<0$), $t(dd\sigma)$ ($>0$), and $t(dd\delta)$ ($>0$) are used.  We therefore have $t_{xy}t_{x^2-y^2}<0$, so that the valence band top and conduction band bottom are located at the same $\bm{k}$-point of the Brillouin Zone.}, 
so that the system has a direct band gap as shown in Fig.~\ref{eics-si-fig2}(a), resulting in a ferro EI state 
with $\bm{Q}=0$ [see Fig.~\ref{eics-si-fig2}(c)].  

We understand in Eq.~(\ref{eics-sa-eq15}) that the spin density in the EI state makes the density wave with modulation $\bm{Q}$, 
but we find in Figs.~\ref{eics-si-fig1} and \ref{eics-si-fig2} that the excitonic SDW state is quite different from the conventional SDW state with an antiferromagnetic order.  
In the EI state in the tight-binding approximation, the charge and spin densities are distributed anisotropically in each unit cell (or atom), 
which may therefore be described suitably by multipole moments, as we will discuss in the next subsection.

\subsection{Multipole moments} \label{eics-mm}
Now, let us describe the EI states in terms of the multipole moments, taking the previous model as an example, 
where the valence $s$ and conduction $p_z$ orbitals are hybridized due to excitonic ordering.  
First, the electric monopole moment, which corresponds to the total charge in each unit cell (or atom), is defined as 
\begin{align}
Q_{0} (\bm{R}_i)  
&\equiv -e \int d\bm{r} \rho_i(\bm{r}) \notag \\
& = -e \sum_{\sigma}\sum_{\alpha,\beta} \left[ \int d\bm{r} \phi_{i\alpha} (\bm{r})\phi_{i\beta} (\bm{r}) \right] \langle c^{\dag}_{i\alpha\sigma}c_{i\beta\sigma} \rangle ,\label{eics-mm-eq1}
\end{align}
where $-e$ is the elementary charge.  Using the orthogonality of the atomic orbitals 
\begin{align}
\int d\bm{r}\phi_{i\alpha} (\bm{r})\phi_{i\beta}(\bm{r}) = \delta_{\alpha,\beta},
\label{eics-mm-eq2}
\end{align}
we find that $Q_0$ in Eq.~(\ref{eics-mm-eq1}) becomes 
\begin{align}
Q_0 (\bm{R}_i) 
= -e \sum_{\alpha} n_{i\alpha} ,
\label{eics-mm-eq3}
\end{align}
given simply as a sum of $n_{i\alpha} $.  
In the same way, the magnetic dipole moment, which corresponds to the integrated magnetization in each unit cell (or atom), is given as 
\begin{align}
\bm{M}_0  (\bm{R}_i) 
 \equiv  -g\mu_{\rm B} \int d\bm{r} \bm{s}_i(\bm{r})  
=  -g\mu_{\rm B}\sum_{\alpha}  \bm{m}_{i\alpha},  
\label{eics-mm-eq4}
\end{align}
where $\bm{M}_0 = (M^x_0, M^y_0, M^z_0)$, $g$ is the Land\'e $g$ factor, and $\mu_{\rm B}$ is the Bohr magneton. 
We find that, in the present example of the $s$ and $p_z$ orbitals, we have the EI state with $n_{i\alpha}=n_{\alpha}$ and $\bm{m}_{i\alpha}=0$, 
so that the electric monopole moment (or total charge) $Q_0$ remains unchanged, $Q_0(\bm{R}_i)  = -e\sum_{\alpha} n_{\alpha}$, 
and the magnetic dipole moment (or net magnetization) $\bm{M}_0$ vanishes, $\bm{M}_0(\bm{R}_i) =0$.  

Next, let us discuss the higher rank multipole moments derived by the excitonic order $\langle c^{\dag}_{ip_z}c_{is} \rangle\ne 0$, 
whereby we evaluate an electric dipole moment in the spin-singlet EI state.  
The electric dipole moment of the $z$ direction is defined as~\cite{ZFB10}
\begin{align}
Q_z (\bm{R}_i)
&\equiv -e \int d\bm{r} z \rho_i(\bm{r}) \notag \\
&= -e \sum_{\sigma}\sum_{\alpha,\beta} \left[ \int d\bm{r} z \phi_{i\alpha} (\bm{r})  \phi_{i\beta} (\bm{r}) \right] \langle c^{\dag}_{i\alpha\sigma}c_{i\beta\sigma} \rangle .  
\label{eics-mm-eq5}
\end{align}
The integral part in Eq.~(\ref{eics-mm-eq5}) becomes 
\begin{align}
&\int d\bm{r} z \phi_{i\alpha} (\bm{r})  \phi_{i\alpha} (\bm{r}) = 0, 
\label{eics-mm-eq6} \\
&\int d\bm{r} z \phi_{is} (\bm{r})  \phi_{ip_z} (\bm{r})  \equiv \Gamma^{sp_z}_{z} \ne 0, 
\label{eics-mm-eq7}
\end{align}
so that we obtain the electric dipole moment as
\begin{align}
Q_z (\bm{R}_i)  = -2e \Gamma^{sp_z}_{z} \Phi_{is} .
\label{eics-mm-eq8}
\end{align}
When the spin-singlet EI state occurs with $\Phi_{is}  \ne 0$ as in Eq.~(\ref{eics-sa-eq8}), this quantity becomes finite as 
\begin{align}
Q_z (\bm{R}_i)  = -2e \Gamma^{sp_z}_{z} \Phi_{s}  \cos \bm{Q}\cdot \bm{R}_i \ne 0 .
\label{eics-mm-eq9}
\end{align}
We also find that the electric dipole moment of the $x$ and $y$ directions vanishes, i.e., $Q_x=Q_y=0$, because the integral part of the wave functions vanishes.  
Therefore, depending on the shape of the wave functions of the valence and conduction bands, only the electric dipole moment of the $z$ direction becomes finite.  

In the same way, the multipole moment for the spin density is given as
\begin{align}
\bm{M}_z (\bm{R}_i) 
\equiv   -g\mu_{\rm B} \int d\bm{r} z \bm{s}_i(\bm{r})  
=  -2 g\mu_{\rm B}  \Gamma^{sp_z}_{z} \bm{\Phi}_{it} , 
\label{eics-mm-eq10}
\end{align}
where $\bm{M}_z = (M^x_z, M^y_z, M^z_z)$.  
The multipole moment in Eq.~(\ref{eics-mm-eq10}) is given as a product of the dipole distribution of the electron density  $\Gamma^{sp_z}_{z}$ 
and the spin polarization (magnetic dipole) $\bm{\Phi}_{it}$, which then results in an on-site magnetic quadrupole~\cite{KA14-1}.  
When the spin-triplet EI state is realized with $\bm{\Phi}_{it}  \ne 0$ as in Eq.~(\ref{eics-sa-eq14}), the magnetic multipole moment $\bm{M}_z$ becomes finite as 
\begin{align}
\bm{M}_z (\bm{R}_i) 
=  -2 g\mu_{\rm B}  \Gamma^{sp_z}_{z} \bm{\Phi}_{t}  \cos \bm{Q}\cdot \bm{R}_i \ne 0. 
\label{eics-mm-eq11}
\end{align}
Likewise, the multipole moments depend on the atomic wave functions of the valence and conduction bands; e.g., 
when the spin-triplet excitonic order occurs with the $d_{xy}$ and $d_{x^2-y^2}$ orbitals, 
the higher rank multipole moment becomes finite, as shown in Fig.~\ref{eics-si-fig2}.  

Let us generalize the present discussion by means of a mapping of the charge and spin densities onto 
the spherical harmonics~\cite{Sc55,Ku08,KKK09,Ku12,CGN10,BCGetal09}. 
The multipole moments for the charge and spin densities may, respectively, be defined as 
\begin{align}
Q_{lm} (\bm{R}_i)
 \equiv -e &\int d\bm{r}  \left[ r^lZ_{lm}(\hat{\bm{r}}) \right] \rho_i(\bm{r}), 
 \label{eics-mm-eq12} \\
\bm{M}_{lm} (\bm{R}_i)
 \equiv -g\mu_{\rm B}  &\int d\bm{r}  \left[ r^lZ_{lm}(\hat{\bm{r}}) \right] \bm{s}_i(\bm{r}) , 
\label{eics-mm-eq13}
\end{align}
where we define $Z_{lm}(\hat{\bm{r}}) \equiv  \sqrt{4\pi/(2l+1)} Y_{lm}(\hat{\bm{r}})$ with the real spherical harmonics $Y_{lm}(\hat{\bm{r}})$ 
(sometimes called the tesseral harmonics).   $\hat{\bm{r}}=(\theta,\varphi)$ indicates the angular coordinates.  
Using the tesseral harmonics, we obtain the multipole moments in Eq.~(\ref{eics-mm-eq12}) as 
$Q_{00}=Q_0$, $Q_{10}=Q_z$, $Q^{(c)}_{11}=Q_x$, $Q^{(s)}_{11}=Q_y$, $\cdots$ (see the Appendix).  
We introduce a vector $\bm{M}_{lm}  = (M^x_{lm},M^y_{lm},M^z_{lm})$ in Eq.~(\ref{eics-mm-eq13}) to indicate the spin direction.  
Defining the integral part of the wave functions 
\begin{align}
\Gamma^{\alpha\beta}_{lm} \equiv  \int d\bm{r}  \left[ r^lZ_{lm}(\hat{\bm{r}}) \right] \phi_{i\alpha} (\bm{r}) \phi_{i\beta} (\bm{r}) 
\label{eics-mm-eq14}
\end{align}
as in Eq.~(\ref{eics-mm-eq7}), we obtain 
\begin{align}
Q_{lm} (\bm{R}_i)
= -e &\sum_{\alpha,\beta} \sum_{\sigma} \Gamma^{\alpha\beta}_{lm} \langle c^{\dag}_{i\alpha\sigma}c_{i\beta\sigma} \rangle, 
\label{eics-mm-eq15} \\
\bm{M}_{lm} (\bm{R}_i)
 =-\frac{g\mu_{\rm B}}{2}  &\sum_{\alpha,\beta}  \sum_{\sigma,\sigma'}  \Gamma^{\alpha\beta}_{lm} \langle c^{\dag}_{i\alpha\sigma}\bm{\sigma}_{\sigma\sigma'}c_{i\beta \sigma'} \rangle, 
\label{eics-mm-eq16}
\end{align}
whereby we find that the EI state is characterized not only by the order parameter $\langle c^{\dag}_{i\alpha}c_{i\beta} \rangle$ ($\alpha \ne \beta$) but also by the integral part $\Gamma^{\alpha\beta}_{lm}$.  
Thus, the higher multipole moments $Q_{lm}$ or $\bm{M}_{lm}$ become finite when $\langle c^{\dag}_{i\alpha}c_{i\beta} \rangle \ne 0$ and $\Gamma^{\alpha\beta}_{lm} \ne 0$.  
Whether $\Gamma^{\alpha\beta}_{lm}$ is finite or not may be estimated from the integral of the product of $Z_{lm}(\hat{\bm{r}})$ and spherical harmonics in the $\alpha$ and $\beta$ orbitals.  
Correspondence between the nonvanishing $\Gamma^{\alpha\beta}_{lm}$ and the orbitals $\alpha$ and $\beta$ is summarized in the Appendix.  

$Q_{lm}$ in Eq.~(\ref{eics-mm-eq12}) corresponds exactly to the electric multipole moment \cite{Sc55,Ku08,ZFB10}, 
where $l$ is the rank of the electric multipole moments called the electric monopole ($l=0$), dipole ($l=1$), quadrupole ($l=2$), octupole ($l=3$), hexadecapole ($l=4$), dotriacontapole ($l=5$), etc.  
Note however that the definition of the multipole moments for the spin density, which is given in Eq.~(\ref{eics-mm-eq13}), is slightly different 
from the definition of the usual magnetic multipole moments~\cite{Sc55,Ku08}.  
Here, we divide the multipole moment in Eq.~(\ref{eics-mm-eq16}) into the integral part of the orbitals and the expectation value of the spin polarization, 
just as in the definition of the multipole moment for the charge density given in Eqs.~(\ref{eics-mm-eq12}) and (\ref{eics-mm-eq15}).  
This classification of the multipole moments we adopted essentially corresponds to the classification made by Cricchio {\it et al.}~\cite{CGN10,BCGetal09}, 
where the uncoupled double tensors are used in a model without spin-orbit coupling.  
In this definition, $\bm{M}_0$ corresponds to the magnetic dipole moment [see Eq.~(\ref{eics-mm-eq4})].  
Thus, we set $l=0$ as the rank-1 magnetic multipole moment and it may be appropriate to call $\bm{M}_{lm}$ the $(l+1)$th rank magnetic multipole moment in this paper.  
The spin density distribution given in Fig.~\ref{eics-si-fig2}, which corresponds to $\bm{M}^{(s)}_{44}=\bm{M}_{xy(x^2-y^2)}\ne 0$, is then called the magnetic dotriacontapole ($l+1=5$). 

Let us emphasize here that, in the spin-singlet (spin-triplet) EI states, when they are derived from the valence and conduction bands composed of the orthogonal orbitals, 
the change in the charge (spin) density distributions occurs within each unit cell (or atom) and the higher-rank electric (magnetic) moments become finite.  
Therefore, the multipole moments, which are ordered with modulation vector $\bm{Q}$, form the complicated charge and spin density waves.  
In this sense, the excitonic CDW and SDW states may be called the electric multipole density-wave state and magnetic multipole density-wave state, respectively.  
We may also point out that the higher-rank electric or magnetic multipole orders caused by the excitonic instability, which are observable in principle, 
might be regarded as hidden orders that are not easy to detect experimentally (see also Sec.~V).

\subsection{Effects of cross hopping} \label{eics-hy}
We have shown in the previous subsections that the net magnetization (or magnetic dipole moment) in the unit cell (or atom) 
does not appear in the spin-triplet EI state when the valence and conduction bands, which are composed of orthogonal atomic orbitals, 
have no hybridization with each other in the normal state.  
However, the SDW states of chromium~\cite{Ri70,So78,Fa88,GVA02} and 
iron-based superconductors~\cite{MSW08,CEE08,HCW08,VVC09,BT09-1,BT09-2,EC10,KECetal10,VVC10,FS10,KEM11,ZTB11,FCKetal12,SBT14}, 
which are sometimes regarded as the spin-triplet excitonic ordering, actually exhibit the antiferromagnetic (or conventional SDW) orderings 
with a nonvanishing net magnetization in each atom (or unit cell).  
How do we reconcile these two facts?  
Here, we will show that the nonvanishing magnetization can appear in each atom (or unit cell) in the spin-triplet EI state 
when the valence and conduction bands come from the atomic orbitals, which are orthogonal in the same site but are nonorthogonal 
with a nonvanishing hopping integral between different orbitals in the neighboring sites (cross hopping).  
This result explains the situation in real materials, where the energy bands are usually constructed by the hybridization 
of many nonorthogonal atomic orbitals.

\begin{figure}[t]
\begin{center}
\includegraphics[width=\columnwidth]{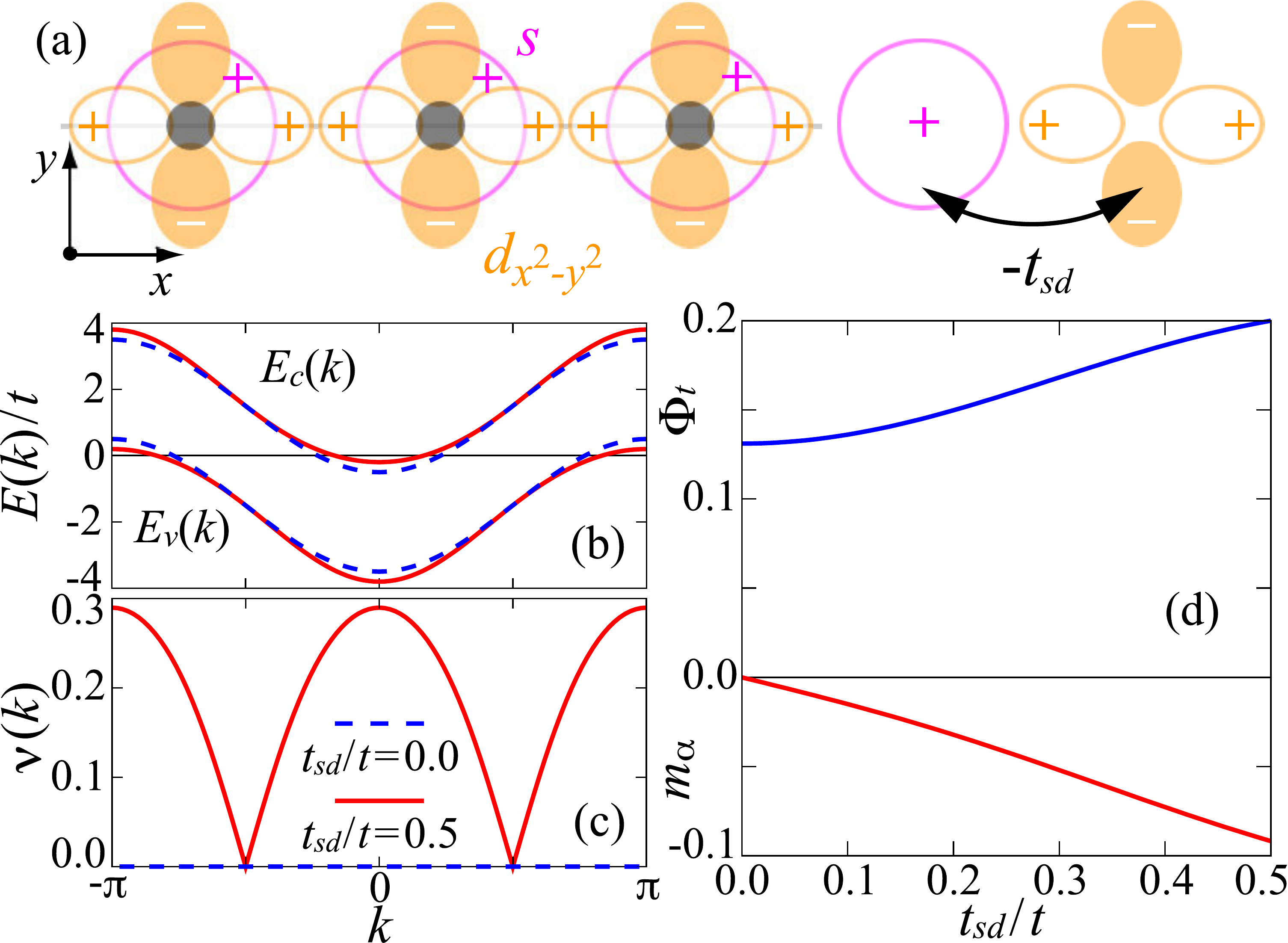}
\caption{(Color online) 
(a) Schematic representation of the one-dimensional chain composed of the valence $s$ and conduction $d_{x^2-y^2}$ orbitals.  
Cross-hopping integral $t_{sd}$ between the neighboring $s$ and $d_{x^2-y^2}$ orbitals is also shown.  
(b) Corresponding noninteracting band dispersions and 
(c) off-diagonal component of the unitary transformation $\nu(k)$, 
where we assume $t_{sd}/t = 0.0$ (dashed line) and $0.5$ (solid line) with $t_s=t_d=t$ and $(\varepsilon_d-\varepsilon_s)/t=3$.  
(d) Spin-triplet orbital diagonal and off-diagonal order parameters as a function of $t_{sd}$ for our model in the Hartree-Fock mean-field approximation, 
where we assume $U_s/t=U_d/t=2$, $U'/t=1$, and $J/t=J'/t=0.5$.  We find $m_{\alpha}=m_s = m_d$ owing to $U_s/t_s=U_d/t_d$.   
}\label{eics-si-fig3}
\end{center}
\end{figure}

Let us assume an example, where the valence and conduction bands are composed of the $s$ and $d_{x^2-y^2}$ orbitals, 
respectively, in the one-dimensional chain [see Fig.~\ref{eics-si-fig3}(a)].  The two orbitals are orthogonal in the same 
site but have the nonvanishing cross hopping $t_{sd}$ between the neighboring sites.  
Hereafter, we sometimes abbreviate $d_{x^2-y^2}$ to $d$ for simplicity.  
The noninteracting tight-binding Hamiltonian of this system is given as 
\begin{align}
\mathcal{H}_{e}  
&=\sum_{\alpha=s,d} \Bigl( \varepsilon_{\alpha}\sum_{i,\sigma} c^{\dag}_{i\alpha\sigma}c_{i\alpha\sigma}  -t_{\alpha}\sum_{\langle i,j\rangle,\sigma} c^{\dag}_{i\alpha\sigma}c_{j\alpha\sigma} \Bigr) 
\notag \\
&\;\;\;\; -t_{sd} \sum_{\langle i,j\rangle,\sigma} \left( c^{\dag}_{is\sigma}c_{jd\sigma} + c^{\dag}_{id\sigma}c_{js\sigma} \right)
\notag \\
&=\sum_{k,\sigma}\left(c^{\dag}_{ks\sigma} \ c^{\dag}_{kd\sigma} \right)
\left(\begin{array}{cc}
\varepsilon_{s}(k) & t_{sd}(k)  \\
t_{sd}(k) & \varepsilon_{d}(k)  \\
\end{array}\right)
\left( \begin{array}{c}
c_{ks\sigma} \\ 
c_{kd\sigma} \\
\end{array}\right), 
\label{eics-vcmo-eq1}
\end{align}
where $\varepsilon_{\alpha}$ is the energy level of the $\alpha$ $(=s,~d)$ orbital and 
$t_{\alpha}$  is the hopping integral between the $\alpha$ orbitals in the neighboring sites. 
The orbital diagonal and off-diagonal components in momentum space are given by 
$\varepsilon_{\alpha}(k) = \varepsilon_{\alpha} - 2t_{\alpha}\cos k$ and $t_{sd}(k) = - 2t_{sd}\cos k$, respectively.  
We assume $\varepsilon_s < \varepsilon_d$, so that the valence (conduction) band includes a large component of the $s$ ($d_{x^2-y^2}$) orbital.  
The diagonalized noninteracting band dispersions are obtained by the unitary transformation 
$\gamma_{k\mu\sigma} = \sum_{\alpha} \zeta_{\mu\alpha}(k,\sigma)c_{k\alpha\sigma}$, 
which connects between the band $\mu$ and orbital $\alpha$.  
The valence band $E_v(k)$ and conduction band $E_c(k)$ are given by 
\begin{align}
E_{v(c)} (k)  = \eta(k) - (+) \sqrt{\xi^2(k) + t^2_{sd}(k)}  
\label{eics-vcmo-eq2}
\end{align}  
with $2\eta(k) = \varepsilon_{d}(k)+\varepsilon_{s}(k)$ and $2\xi(k) = \varepsilon_{d}(k)-\varepsilon_{s}(k)$. 
The unitary transformation connecting between the band $\mu$ (=$v$, $c$) and orbital $\alpha$ (=$s$, $d$) 
is given by 
\begin{align}
\left( \begin{array}{c}
\gamma_{kv\sigma} \\ 
\gamma_{kc\sigma} \\
\end{array}\right)
=
\left(\begin{array}{cc}
\sqrt{1-\nu^2(k)} & -\nu(k)  \\
\nu(k) &  \sqrt{1-\nu^2(k)}  \\
\end{array}\right)
\left( \begin{array}{c}
c_{ks\sigma} \\ 
c_{kd\sigma} \\
\end{array}\right), 
\label{eics-vcmo-eq3}
\end{align}
where the off-diagonal component $\nu(k)$ is given by 
\begin{align}
\nu^2(k) = \frac{1}{2} \left( 1 - \frac{\xi(k)}{\sqrt{\xi^2(k)+t^2_{sd}(k)}} \right). 
\label{eics-vcmo-eq4}
\end{align} 
$\nu(k)$ indicates the weight of the $s$ ($d_{x^2-y^2}$) orbital component in the conduction (valence) band.  
The band dispersions and $\nu(k)$ are shown, respectively, in Figs.~\ref{eics-si-fig3}(b) and \ref{eics-si-fig3}(c), 
where we assume $t_{sd}=0$ or $0.5t$ with $t_s=t_d=t$ and $(\varepsilon_d-\varepsilon_s)=3t$.  
We find that, irrespective of $t_{sd}$, the valence band top and conduction band bottom are located at $k=\pm \pi$ and $k=0$, respectively, 
so that this system may have an excitonic instability with modulation $Q=\pi$.  
At $t_{sd}=0$, where $\nu(k)=0$, the valence (conduction) band comes purely from the $s$ ($d_{x^2-y^2}$) orbital.  
At $t_{sd}=0.5t$, where $\nu(k)$ has a large value around $k=0$ and $\pm\pi$ but it vanishes at $k=\pm\pi/2$, reflecting $t_{sd}(k)= -2t_{sd}\cos (k)$, 
the conduction band bottom at $k=0$ (valence band top at $k=\pm\pi$) acquires the $s$ ($d_{x^2-y^2}$) orbital component.  
Therefore, at $t_{sd}>0$, the valence band around $k=\pm\pi$ and conduction band around $k=0$ include the same orbital components, 
so that the intraorbital Coulomb interaction may have an impact on the bands in these regions.  

\begin{figure}[t]
\begin{center}
\includegraphics[width=\columnwidth]{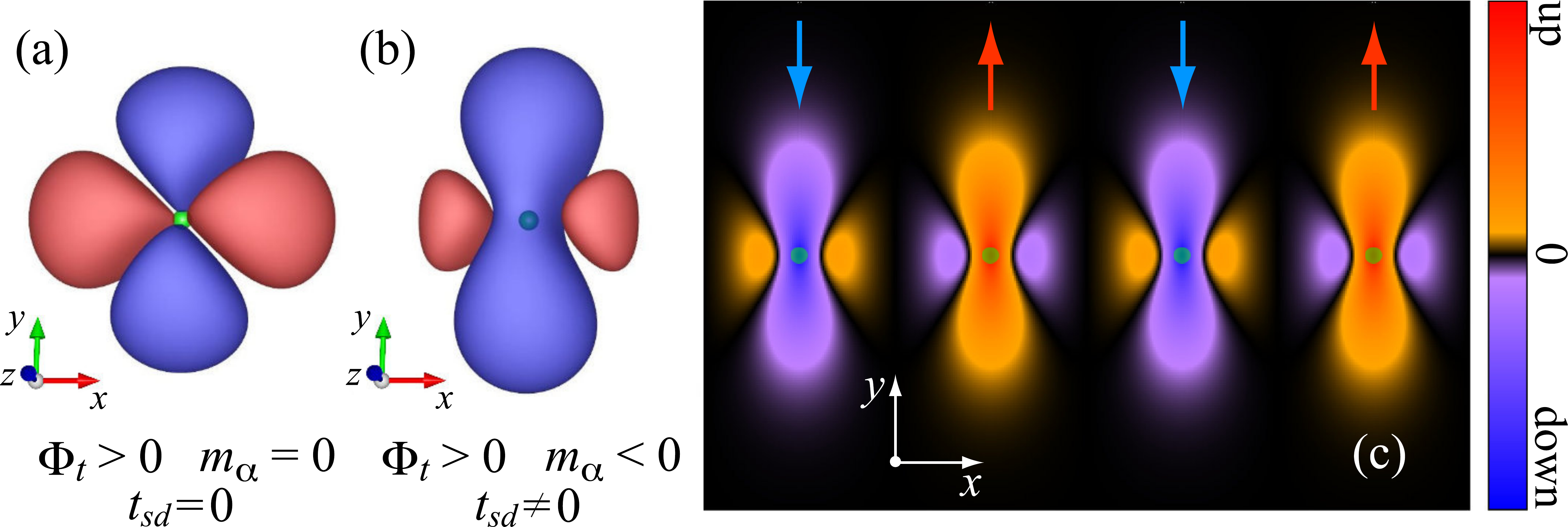}
\caption{(Color online) 
Isosurface of the spin density in the unit cell at $R_i=0$ when 
(a) $\Phi_{t}>0$ and $m_{\alpha}=0$ at $t_{sd}=0$ and 
(b) $\Phi_{t}>0$ and $m_{\alpha}<0$ at $t_{sd}\ne 0$.  
(c) Spin density distribution of our one-dimensional model with $\Phi_{t}>0$ and $m_{\alpha}<0$ at $t_{sd}\ne 0$, 
where positive (up-spin) and negative (down-spin) parts of the spin density are indicated by red and blue, respectively.  
As in Fig.~\ref{eics-si-fig1}, the radial wave function of the $1s$ orbital is slightly broadened to exaggerate 
the character of the spin density distributions although the exact spherical harmonics is assumed for the angular 
dependencies of the $s$ and $d_{x^2-y^2}$ orbitals.  
}\label{eics-si-fig4}
\end{center}
\end{figure}

To investigate the magnetic structure of the corresponding two-band Hubbard model with this band dispersion, 
we apply the Hartree-Fock mean-field approximation, taking into account the intraorbital Coulomb ($U_{s}$, $U_{d}$), 
interorbital direct Coulomb ($U'$), interorbital exchange ($J$), and pair-hopping ($J'$) interactions.  
We then obtain the orbital diagonal and off-diagonal order parameters in the $z$ direction 
$m^z_{\alpha} = \sum_{k,\sigma} \sigma \langle c^{\dag}_{k+Q\alpha\sigma}c_{k\alpha\sigma}\rangle /2N$ and 
$\Phi^z_{t} = \sum_{k,\sigma} \sigma \langle c^{\dag}_{k+Q d \sigma}c_{k s \sigma}\rangle/2N$ 
by solving the self-consistent equations, where we extend the $2\times 2$ matrix in Eq.~(\ref{eics-vcmo-eq1}) to a $4 \times 4$ matrix.  
The calculated results for $\Phi^z_t$ and $m^z_{\alpha}$ are shown in Fig.~\ref{eics-si-fig3}(d) as a function of $t_{sd}$, 
where we assume $t_s=t_d=t$, $(\varepsilon_d-\varepsilon_s)/t=3$, $U_s/t=U_d/t=2$, $U'/t=1$, and $J/t=J'/t=0.5$.  
At $t_{sd}=0$, we obtain the solution with $\Phi^z_{t}>0$ and $m^z_{\alpha}=0$, which is consistent with the results shown in Secs.~\ref{eics-sa} and \ref{eics-mm}.  
At $t_{sd}\ne 0$, however, we find the solution with a nonvanishing magnetization $m_{\alpha}\ne 0$, where $\Phi^z_t$ and $m^z_{\alpha}$ are enhanced with increasing $t_{sd}$.  
Therefore, when the valence and conduction bands include the same orbital component due to the cross hopping $t_{sd}$, 
the net magnetization (magnetic dipole moment) $M^z_{0} = -g\mu_{\rm B}\sum_{\alpha} m^z_{\alpha}$ appears in each unit cell (or atom), 
just as in the conventional SDW states.  

The excitonic SDW state with nonvanishing magnetization in each unit cell (or atom) manifests itself in the illustration of the spin density distribution in real space.  
In Figs.~\ref{eics-si-fig4}(a) and \ref{eics-si-fig4}(b), we illustrate the spin density in the $i$-th unit cell (or atom) 
$s^z_i(\bm{r})  = \left[\phi^{2}_{is} (\bm{r}) m^z_{s} + \phi^{2}_{id} (\bm{r}) m^z_{d} + 2 \phi_{is} (\bm{r}) \phi_{id} (\bm{r}) \Phi^z_t \right] \cos Q R_i$ 
with $m^z_{\alpha}=0$ and $m^z_{\alpha}\ne 0$, respectively.  
At $t_{sd}=0$, we obtain the solution with $m^z_{\alpha}=0$ and $\Phi^z_t \ne 0$, which indicates that the magnetic octupole moment appears 
but the magnetic dipole moment is absent, as shown in Fig.~\ref{eics-si-fig4}(a), in accordance with the discussion in Sec.~\ref{eics-sa}.  
At $t_{sd}\ne 0$, however, we obtain the solution with $m^z_{\alpha}\ne 0$ and $\Phi^z_t \ne 0$, which indicates that 
the negative (down-spin) part is enhanced along the $y$ axis and the positive (up-spin) part is reduced along the $x$ axis, 
resulting in the nonvanishing net magnetization, as shown in Fig.~\ref{eics-si-fig4}(b).  
Thus, the magnetic dipole and octupole moments concurrently appear in each unit cell (or atom).  
In the one-dimensional system with $Q=\pi$, the spin polarization inverts alternately over the unit cells (or atoms), as shown in Fig.~\ref{eics-si-fig4}(c).  
However, in contrast to the conventional SDW state, where the spin density in an atom aligns in the same direction as in a single-orbital Hubbard system, 
the excitonic SDW state with nonvanishing magnetization realized in multiorbital systems contains a spatial structure of the spin density within 
the unit cell (or atom), reflecting the higher-rank multipole moments.  

As we have shown here, the net magnetization (or integrated local dipole moment) in each unit cell (or atom) 
$\bm{M}_{0} = -g\mu_{\rm B}\sum_{\alpha} \bm{m}_{\alpha} $ appears when the valence and conduction 
bands include the same orbital component due to the cross hopping. 
This may explain why the SDW states appear in chromium and iron-based superconductors (if they are of the excitonic origin).  
We have to note however that, in contrast to the conventional SDW state as in the single-orbital system, 
nonvanishing higher-rank multipole moments are superimposed on the SDW states, 
which can be seen in the local spin density $\bm{s}_{i}(\bm{r})$ but cannot be seen in $\bm{m}_{i\alpha}$.  
In the same way, a conventional CDW modulation may be superimposed on the excitonic CDW state 
when the valence and conduction bands include the same orbital component due to the cross hopping.

\section{Multiorbitals in Different Atoms}
Next, let us consider the case where the valence and conduction bands are composed of orbitals in  {\it different} atoms, 
which is thought to occur in some candidate materials such as 1$T$-TiSe$_2$ and Ta$_2$NiSe$_5$ \cite{ZF77,FGH97,KTKetal13}.  
When there are several atoms in a unit cell, we have to consider the spatial position of the $\alpha$ orbital, $\bm{r}_{\alpha}$. 
The Bloch function for the $\alpha$ orbital is given in the tight-binding approximation as 
\begin{align}
\psi_{\bm{k}\alpha}(\bm{r})
 = \frac{1}{\sqrt{N}}\sum_{i} e^{i\bm{k}\cdot \bm{R}_i} \phi_{\alpha} (\bm{r}-\bm{r}_{\alpha}-\bm{R}_i).  
\label{eics-da-eq0}
\end{align} 
The field operator in real space is then given by 
\begin{align}
\Psi_{\sigma}(\bm{r})
 = \sum_{i}\sum_{\alpha} \phi_{\alpha} (\bm{r}-\bm{r}_\alpha-\bm{R}_i)c_{i\alpha\sigma}, 
 \label{eics-da-eq1}
\end{align} 
whereby we can evaluate the charge and spin densities of EI using Eqs.~(\ref{ei-eq6}) and (\ref{ei-eq7}). 

Here, we assume a two-band system for simplicity, where the valence ($a$) and conduction ($b$) bands composed, 
respectively, of the $a$ and $b$ orbitals located in different atoms in a unit cell $i$.  
The field operator in the $i$-th unit cell, which contains the orbitals $a$ and $b$, is then given as 
\begin{align}
\Psi_{i\sigma}(\bm{r}) =  \phi_{ia} (\bm{r}-\bm{r}_a)c_{ia\sigma} + \phi_{ib} (\bm{r}-\bm{r}_b)c_{ib\sigma},  
\label{eics-da-eq2}
\end{align}
where we write $\phi_{i\alpha} (\bm{r}-\bm{r}_{\alpha}) =  \phi_{\alpha} (\bm{r}-\bm{r}_{\alpha}-\bm{R}_i) $ for simplicity. 
Using Eq.~(\ref{eics-da-eq2}), we obtain the density of electrons of spin $\sigma$, 
$\rho_{i\sigma}(\bm{r}) =  \langle \Psi^{\dag}_{i\sigma}(\bm{r}) \Psi_{i\sigma}(\bm{r}) \rangle $, as 
\begin{align}
\rho_{i\sigma}(\bm{r}) 
&=  \phi^{2}_{ia} (\bm{r}-\bm{r}_a)\langle c^{\dag}_{ia\sigma}c_{ia\sigma}\rangle 
+ \phi^{2}_{ib} (\bm{r}-\bm{r}_b)\langle c^{\dag}_{ib\sigma}c_{ib\sigma}\rangle \notag \\
&+ \phi_{ia} (\bm{r}-\bm{r}_a) \phi_{ib} (\bm{r}-\bm{r}_b)
\left[ \langle c^{\dag}_{ib\sigma}c_{ia\sigma}\rangle 
     + \langle c^{\dag}_{ia\sigma}c_{ib\sigma}\rangle \right],   
\label{eics-da-eq3}
\end{align} 
which leads to the charge density 
$\rho_i(\bm{r}) = \rho_{i\uparrow}(\bm{r}) + \rho_{i\downarrow}(\bm{r})$ and spin density of the $z$ direction 
$2s^z_i(\bm{r}) =  \rho_{i\uparrow}(\bm{r}) - \rho_{i\downarrow}(\bm{r})$.  

The change in the electronic density distribution due to the excitonic ordering $\langle c^{\dag}_{ib\sigma}c_{ia\sigma}\rangle \ne 0$ 
is given by the third term of Eq.~(\ref{eics-da-eq3}), which indicates the change in the electron density between the two orbitals $a$ and $b$.  
When the excitonic density wave is given as 
$\langle c^{\dag}_{ib\sigma}c_{ia\sigma}\rangle = \langle c^{\dag}_{ia\sigma}c_{ib\sigma}\rangle = A_{\sigma} \cos \bm{Q}\cdot\bm{R}_i + C_{\sigma}$,  
the density of electrons between the two atoms is enhanced in a unit cell, forming a bonding orbital, but it is reduced in the neighboring unit cells, forming an antibonding orbital.  
Therefore, when the valence and conduction bands are composed of orbitals in different atoms, the excitonic ordering is nothing but a bond order formation; 
in the present case, it is the formation of the bond order wave of spatial modulation $\bm{Q}$.  
Note that the third term in Eq.~(\ref{eics-da-eq3}) includes the product of the wave functions $\phi_{ia} (\bm{r}-\bm{r}_a) \phi_{ib} (\bm{r}-\bm{r}_b)$, 
so that the electron distribution depends on the sign of these wave functions; if $\langle c^{\dag}_{ib\sigma}c_{ia\sigma}\rangle >0$ due to excitonic ordering, 
the bonding (antibonding) orbital is formed in the positive (negative) part of the product of the wave function $\phi_{ia} (\bm{r}-\bm{r}_a) \phi_{ib} (\bm{r}-\bm{r}_b)$.  

In contrast to the EIs formed in a single atom, the interorbital exchange interactions, such as Hund's rule coupling, is weak between different atoms, 
so that the spin-triplet excitonic state is unlikely to be stabilized~\cite{HR68-2,KZFetal15}.  However, the density of electrons between atoms 
is modulated due to the spin-singlet excitonic ordering, which necessarily influences the position of the neighboring atoms due to Coulomb interactions.  
We have shown that, when the valence and conduction bands are composed of orbitals in different atoms, the spin-singlet excitonic state 
is most likely to be stabilized with the help of the lattice distortion of the system~\cite{KZFetal15}.  We argue that this state actually occurs in 1$T$-TiSe$_2$ and Ta$_2$NiSe$_5$~\cite{WNS10,MBCetal11,ZFBetal13,KTKetal13}.  

\begin{figure}[!t]
\begin{center}
\includegraphics[width=0.85\columnwidth]{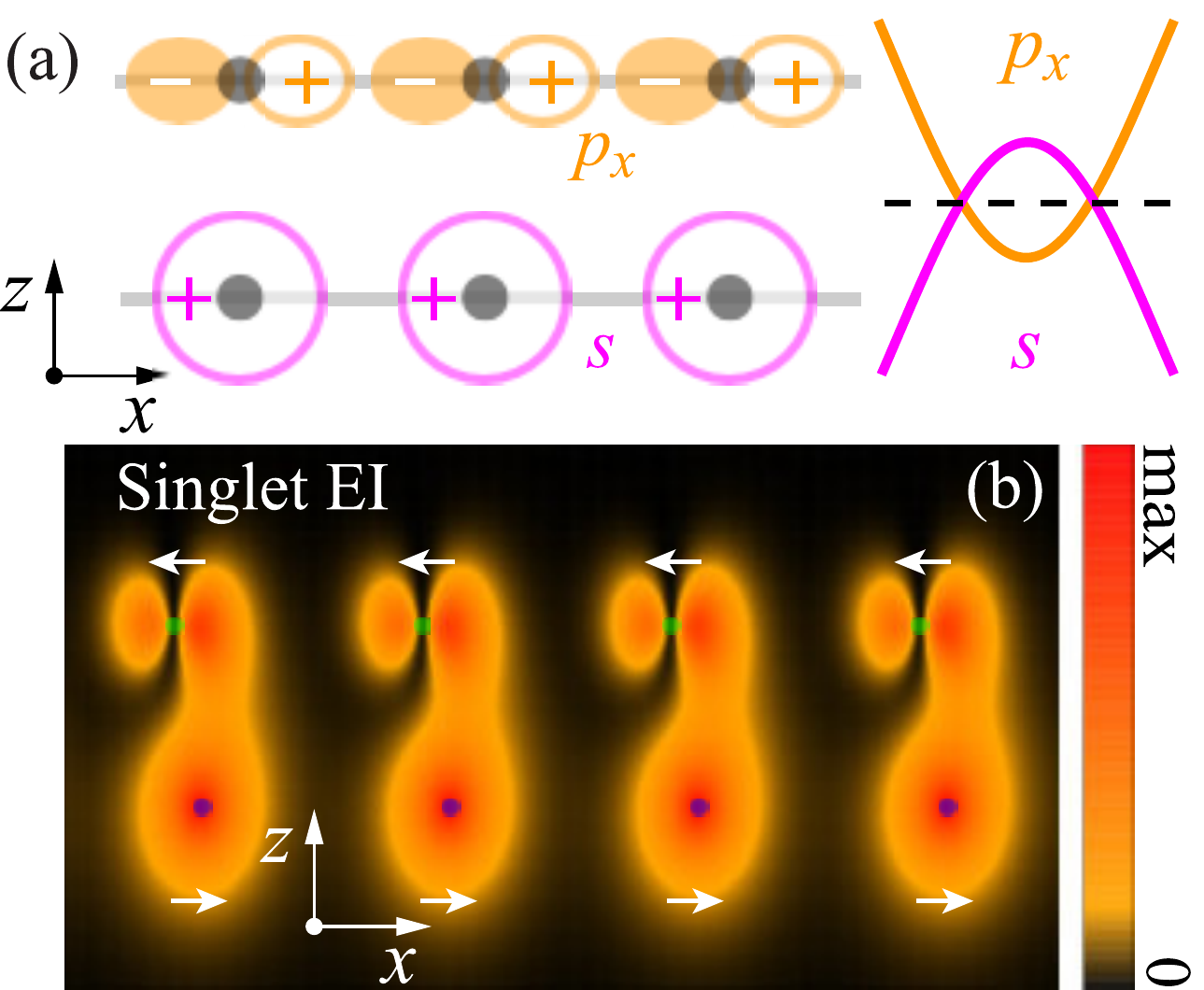}
\caption{(Color online)
(a) Schematic representation of the one-dimensional model with the valence ($s$) and conduction ($p_x$) orbitals in different atoms.  
(b) Charge density distribution of the spin-singlet EI state with $\bm{Q}=0$.  The shear distortion of the lattice is indicated by arrows. 
}\label{eics-di-fig1}
\end{center}
\end{figure}

For an intuitive understanding, we show an example in Fig.~\ref{eics-di-fig1}, where the valence and conduction bands are composed of the $s$ and $p_x$ orbitals in different atoms.  
The unit cells, each of which contains two atoms forming the valence and conduction bands, are arranged in the $x$ direction as a one-dimensional model.  
In this example, the top of the valence band and bottom of the conduction band are located at the same $\bm{k}$ point, so that we have the difference $\bm{Q}=0$.  
The charge density distribution of the spin-singlet EI state in this example is shown in Fig.~\ref{eics-di-fig1}(b).  
We find that the charge density is enhanced (suppressed) in the region where 
$\phi_{ia} (\bm{r}-\bm{r}_a) \phi_{ib} (\bm{r}-\bm{r}_b)\langle c^{\dag}_{ib\sigma}c_{ia\sigma}\rangle>0$ ($<0$).  
Corresponding to the charge density distribution, the lattice may be deformed as shown by arrows in Fig.~\ref{eics-di-fig1}(b).  
The spin-singlet excitonic ordering thus results in the bond order formation.

\section{Discussion and conclusions}

Finally, let us discuss the implications of our results in the materials aspects.  
The condensation of spin-triplet excitons was recently predicted to occur in the proximity of the spin state transition of cobalt oxides~\cite{KA14-2,Ku14,INTetal16,ILTetal16,SK16,TMNetal16}.  
In this system, the valence and conduction bands are formed by the $t_{2g}$ and $e_g$ orbitals, respectively, in the (same) cobalt atoms.
If the EI state is stabilized in this system, the spin-triplet excitonic pairing is favored due to strong Hund's rule coupling in a single atom, 
and therefore the magnetic multipole may appear as the spin-triplet orbital-off-diagonal order, which may have a vanishing magnetic (dipole) moment per site~\cite{KA14-2}.  
Kune\v{s} and Augustinsk\'{y}~\cite{KA14-2} suggested that this type of magnetic multipole order occurs in Pr$_{0.5}$Ca$_{0.5}$CoO$_3$ and 
that the experimental data can be explained comprehensively by the spin-triplet exciton condensation scenario. 
The SDW states observed in chromium metal \cite{So78,GVA02} and iron-based superconductors 
\cite{MSW08,CEE08,HCW08,VVC09,BT09-1,BT09-2,EC10,KECetal10,VVC10,FS10,KEM11,ZTB11,FCKetal12,SBT14} 
have sometimes been argued to be of the excitonic origin as well.  
The valence and conduction bands in these systems are mainly formed by the $3d$ orbitals of the (same) transition-metal atoms, just as in the cobalt oxides.  
However, we note that the SDW states of chromium and iron-based superconductors were studied as the {\it band} off-diagonal order, 
which is in contrast to the cobalt oxide, where the spin-triplet EI state was studied as the {\it orbital} off-diagonal order.  
The band off-diagonal orders are based on the diagonalized noninteracting bands including components of many relevant orbitals, 
and the Coulomb interactions are added as the effective interband interactions in the diagonalized bands around the Fermi levels. 
Here, an electron in the conduction band and a hole in the valence band are quasiparticles formed by the hybridization of many orbitals.  
The total magnetic moment per atom can be finite as discussed in Sec.~\ref{eics-hy} because the conduction and valence bands 
here include the same orbital component.  However, in contrast to the conventional SDW state in single-orbital systems, 
the complicated SDW states with magnetic multipole moments can be realized in such multiorbital systems.  
Cricchio {\it et al.} suggested from first-principles calculations that this type of order actually occurs in iron-based superconductors~\cite{CGN10}. 
In these systems, an unexpectedly low magnetic moment is observed experimentally \cite{KA14-2,CGN10} unlike in the conventional SDW state, which is thought to be due to the vanishing total magnetic (dipole) moment in the presence of higher-rank magnetic multipole moments.  We note that even though the total magnetic moment per site is zero the magnetic multipoles have the local magnetic polarization in each unit cell (or atom), which may be observed by, e.g., resonant x-ray scattering experiments~\cite{KKK09,Ku12,CGN10,FSMetal10}.

The condensation of spin-singlet excitons was recently predicted in 1$T$-TiSe$_2$ and Ta$_2$NiSe$_5$.  
In these systems, the valence and conduction bands are formed by the orbitals located in different atoms; 
in $1T$-TiSe$_2$, the $4p$ orbitals of Se atoms account for the valence bands and the $3d$ orbitals of Ti atoms account for 
the conduction bands~\cite{CMCetal07,MCCetal09,MSGetal10,WNS10,MBCetal11,MMAetal12,ZFBetal13,MMHetal15,WSY15}.  
In Ta$_2$NiSe$_5$, which is a small direct-gap semiconductor, the $3d$ orbitals of Ni atoms form the valence bands and the $5d$ orbitals of Ta atoms 
form the conduction bands~\cite{WSTetal09,WSTetal12,KTKetal13,SWKetal14}. 
If the spin-singlet EI state is realized in these systems, the orbitals located in different atoms are hybridized spontaneously to make the bonding (or antibonding) state, 
which is the bond order formation as we have shown in Sec.~IV.   
The lattice degrees of freedom necessarily couple with the bond order formation in these systems; 
in fact, the lattice distortions corresponding to the vector $\bm{Q}\ne 0$ in TiSe$_2$ and 
the shear lattice distortion corresponding to $\bm{Q}=0$ in Ta$_2$NiSe$_5$ have been observed.  
In 1$T$-TiSe$_2$ and Ta$_2$NiSe$_5$, the electron-phonon coupling and interband Coulomb interaction work cooperatively to stabilize the spin-singlet bond orders.  

We may also point out that the electronic ferroelectricity and antiferroelectricity are the relevant physics in the EI state formation.  
We may actually predict that such states should occur when the even and odd parity orbitals are hybridized spontaneously and 
the electronic density distribution breaks the space-inversion symmetry in each unit cell.  
An example is shown in Fig.~\ref{eics-si-fig1}(d), which was discussed in Sec.~\ref{eics-sa}.  
In fact, the electronic ferroelectricity derived from the hybridization between different-parity orbitals was discussed using the spinless extended Falicov-Kimball model~\cite{POS96-2,Ba02,ZFB10}.  
We moreover note that the spin density distribution of the spin-triplet EI state shown in Fig.~\ref{eics-si-fig1}(f) breaks not only the space-inversion symmetry 
but also the time-reversal symmetry in each unit cell.  In this type of the electronic spin density distributions, one may expect the magnetoelectric effects to occur~\cite{EMS06}. 
In general, when the local wave functions of the valence and conduction bands have different parities, as in the $s$-$p$, $p$-$d$, and $d$-$f$ orbitals, 
the spin-triplet excitonic orders give rise to the magnetic quadrupole, hexadecapole, or tetrahexacontapole orders, which break both the space-inversion and time-reversal symmetries 
in each unit cell.  Thus, one may expect the magnetoelectric effects to occur in such cases.  

To conclude, we have evaluated the charge and spin densities of the spin-singlet and spin-triplet EI states from the local wave functions in the tight-binding approximation.  
We first discussed the charge and spin density distributions of the EI states when the valence and conduction bands are composed of orthogonal orbitals in a single atom.  
We found that the anisotropic charge or spin density distribution occurs in each unit cell (or atom) and higher rank electric or magnetic multipole moment becomes finite, 
depending on the wave functions of the valence and conduction bands, which is the multipole order formation.  
In contrast to the conventional CDW or SDW state, the modulation of the total charge (electric monopole moment) or the net magnetization (magnetic dipole moment) in the unit cell (or atom) 
does not appear when the two orthogonal orbitals are hybridized via a spin-singlet or spin-triplet excitonic ordering.  
However, in the real materials, the energy bands are constructed by the hybridization of many orbitals. 
We then found that the net magnetization in each unit cell (or atom) can appear as in the conventional SDW state if both the conduction and valence bands include the same orbital component.  
We also discussed the electron density distribution in the EI state when the valence and conduction bands are composed of orbitals in different atoms.  
We found that the excitonic ordering enhances (suppresses) the electronic density between atoms when the bonding (antibonding) state is formed between the two orbitals, 
which is therefore nothing but the bond order formation.  

We have studied the simplest two-orbital model throughout this paper.  In real materials, however, 
we may encounter the situation where the relevant bands are composed of more than two orbitals.  
The spin-orbit coupling, which was completely neglected in this paper, can also be important in some 
situations.  We want to leave these issues for future research. 

\begin{acknowledgments}
The authors would like to thank S. Miyakoshi, H. Nishida, and K. Sugimoto for enlightening discussions. 
T.~K.~acknowledges support from a JSPS Research Fellowship for Young Scientists.  
This work was supported in part by a Grant-in-Aid for Scientific Research from JSPS (No. 26400349) of Japan.  
\end{acknowledgments}

\begin{appendix}

\section{Multipole Expansion of Electronic Density in Excitonic Insulators} \label{app-a}
In this appendix, we present the multipole expansion of the electronic density distribution in the EI states 
when the valence and conduction bands are composed of the orbitals in a single atom.  
The field operator for a multiorbital system of a single atom is given by
\begin{align}
\Psi(\bm{r}) = \sum_{\alpha} \phi_{\alpha} (\bm{r})c_{\alpha} , \:\:\: 
\Psi^{\dag}(\bm{r}) = \sum_{\alpha} \phi^{*}_{\alpha} (\bm{r})c^{\dag}_{\alpha},  
\label{eicsmp-eq1}
\end{align} 
where $\phi_{\alpha}(\bm{r})$ is the atomic wave function and $c_{\alpha}$ ($c^{\dag}_{\alpha}$) is 
the annihilation (creation) operator of an electron in the $\alpha$ orbital~\cite{Ba00,KA14-1}.  
Hereafter, we omit the site and spin indices for simplicity because we consider only the atomic wave functions in a single atom.  
We also do not write the elementary charge $e$ explicitly.  The wave function of the $\alpha$ orbital is given by 
\begin{align}
\phi_{\alpha}(\bm{r}) = \phi_{n_{\alpha}l_{\alpha}m_{\alpha}}(\bm{r}) = R_{n_{\alpha}l_{\alpha}}(r)Y_{l_{\alpha}m_{\alpha}}(\hat{\bm{r}}),  
\label{eicsmp-eq2}
\end{align}
where $R_{nl}(r)$ is the radial wave function and $Y_{lm}(\hat{\bm{r}})=Y_{lm}(\theta,\varphi)$ is the spherical harmonics.  
$n_{\alpha}$, $l_{\alpha}$, and $m_{\alpha}$ are the principal,  azimuthal, magnetic quantum numbers of the $\alpha$ orbital, respectively.  
Using Eq.~(\ref{eicsmp-eq1}), we write the electronic density as 
\begin{align}
\rho(\bm{r}) \equiv \langle \Psi^{\dag}(\bm{r}) \Psi(\bm{r}) \rangle = \sum_{\alpha,\beta} \phi^{*}_{\alpha} (\bm{r})\phi_{\beta} (\bm{r}) \langle c^{\dag}_{\alpha}c_{\beta} \rangle, 
\label{eicsmp-eq3}
\end{align}
whereby we find the modification of the electronic density $\rho(\bm{r})$ caused by the spontaneous hybridization 
$\langle c^{\dag}_{\alpha}c_{\beta} \rangle\ne 0$ between orbitals $\alpha$ and $\beta$ due to excitonic ordering.  

Let us describe the character of the EI in terms of the multipole moments, which are defined by the projection of $\rho(\bm{r})$ onto the spherical harmonics~\cite{Ku08,KKK09,Ku12} as 
\begin{align}
Q_{lm} \equiv \int d\bm{r}  [ r^lZ^{*}_{lm}(\hat{\bm{r}}) ] \rho(\bm{r}), 
\label{eicsmp-eq4}
\end{align}
where $Z_{lm}(\hat{\bm{r}}) \equiv  \sqrt{4\pi/(2l+1)} Y_{lm}(\hat{\bm{r}})$ and $Z^{*}_{lm}(\hat{\bm{r}}) = (-1)^mZ^{}_{l-m}(\hat{\bm{r}})$.  
$l$ is the rank of the multipole moments, which are called 
the monopole ($l=0$), 
dipole ($l=1$), 
quadrupole ($l=2$), 
octupole ($l=3$), 
hexadecapole ($l=4$), 
dotriacontapole ($l=5$), etc.  
Using Eq.~(\ref{eicsmp-eq3}), we write the multipole moment as 
\begin{align}
Q_{lm} 
& =  \sum_{\alpha,\beta} \left[ \int d\bm{r}  \phi^{*}_{\alpha} (\bm{r}) r^lZ^{*}_{lm}(\hat{\bm{r}}) \phi_{\beta} (\bm{r}) \right] \langle c^{\dag}_{\alpha}c_{\beta} \rangle \notag \\
& = \sum_{\alpha,\beta}  \Gamma^{\alpha\beta}_{lm} \langle c^{\dag}_{\alpha}c_{\beta} \rangle , 
\label{eicsmp-eq5}
\end{align}
where we define the integral part as 
\begin{align}
\Gamma^{\alpha\beta}_{lm} 
\equiv \int d\bm{r}  \phi^{*}_{\alpha} (\bm{r}) r^lZ^{*}_{lm}(\hat{\bm{r}}) \phi_{\beta} (\bm{r}) . 
\label{eicsmp-eq6}
\end{align}
From Eq.~(\ref{eicsmp-eq5}), we find that the multipole moment is finite, $Q_{lm}\ne 0$,  
when both $\Gamma^{\alpha\beta}_{lm} $ and $ \langle c^{\dag}_{\alpha}c_{\beta} \rangle$ are nonzero. 
A finite value of $ \langle c^{\dag}_{\alpha}c_{\beta} \rangle$ is obtained for a symmetry-broken solution, 
but the value of $\Gamma^{\alpha\beta}_{lm}$ depends on the character of orbitals in the valence and conduction bands. 

Now, let us calculate $\Gamma^{\alpha\beta}_{lm}$ in detail and clarify which rank of the multipole moments is finite 
depending on which orbitals are hybridized.  Using Eq.~(\ref{eicsmp-eq2}), we find 
\begin{align}
\Gamma^{\alpha\beta}_{lm} 
&= \left[ \int r^{2} dr R_{n_{\alpha}l_{\alpha}}(r) r^{l} R_{n_{\beta}l_{\beta}}(r)\right] 
\notag \\
&\times \left[ \int d\Omega Y^{*}_{l_{\alpha}m_{\alpha}}(\hat{\bm{r}}) Z^{*}_{lm}(\hat{\bm{r}}) Y_{l_{\beta}m_{\beta}}(\hat{\bm{r}}) \right], 
\label{eicsmp-eq7}
\end{align}
where $d\Omega = \sin \theta d\theta d\varphi$.  Defining 
\begin{align}
& \Lambda_l(n_{\alpha}l_{\alpha},n_{\beta}l_{\beta}) \equiv \int r^{2} dr R_{n_{\alpha}l_{\alpha}}(r) r^{l} R_{n_{\beta}l_{\beta}}(r),  
\label{eicsmp-eq8} \\
& \Theta_{lm} (l_{\alpha}m_{\alpha},l_{\beta}m_{\beta})  \equiv \int d\Omega Y^{*}_{l_{\alpha}m_{\alpha}}(\hat{\bm{r}}) Z^{*}_{lm}(\hat{\bm{r}}) Y_{l_{\beta}m_{\beta}}(\hat{\bm{r}}) , 
\label{eicsmp-eq9} 
\end{align}
we can divide $\Gamma_{lm}^{\alpha\beta} $ into the radial and angular parts as 
\begin{align}
\Gamma_{lm}^{\alpha\beta} = \Lambda_l(n_{\alpha}l_{\alpha},n_{\beta}l_{\beta})\Theta_{lm} (l_{\alpha}m_{\alpha},l_{\beta}m_{\beta}). 
\label{eicsmp-eq10}
\end{align}
Thus, we can classify $\Gamma_{lm}^{\alpha\beta} $ in terms of $\Theta_{lm} (l_{\alpha}m_{\alpha},l_{\beta}m_{\beta})$ 
because the radial part $\Lambda_l(n_{\alpha}l_{\alpha},n_{\beta}l_{\beta})$ is always nonzero.  

\begin{table}[t]
\begin{center}
{\renewcommand\arraystretch{1.2}
\begin{tabular}{|c|c|c|c|}
\hline
$l$  & $m$ & \makebox[3.5cm]{$(c)$ and $0$} &  \makebox[3.5cm]{$(s)$} 
\\ \hline
0 & 0 & 1 & --
\\ \hline
1 & 0 & $z$ &  \\
  & 1 & $x$ & $y$ 
\\ \hline
2 & 0 & $3z^2-r^2$ & --   \\
  & 1 & $zx$       & $yz$ \\
  & 2 & $x^2-y^2$  & $xy$ \\
\hline
3 & 0 & $z(5z^2-3r^2)$ & --            \\
  & 1 & $x(5z^2-r^2)$  & $y(5z^2-r^2)$ \\
  & 2 & $z(x^2-y^2)$   & $xyz$         \\
  & 3 & $x(x^2-3y^2)$  & $y(3x^2-y^2)$ \\ 
\hline
4 & 0 & $35z^4-30z^2r^2+3r^4$ & --              \\
  & 1 & $zx(7z^2-3r^2)$       & $yz(7z^2-3r^2)$ \\
  & 2 & $(x^2-y^2)(7z^2-r^2)$ & $xy(7z^2-r^2)$  \\
  & 3 & $zx(x^2-3y^2)$        & $yz(3x^2-y^2)$  \\ 
  & 4 & $x^4-6x^2y^2+y^4$     & $xy(x^2-y^2)$   \\ 
\hline
\end{tabular}}
\caption{Correspondence between the tesseral representation and Cartesian coordinate representation.}
\label{eicsmp-table1}
\end{center}
\end{table}

Using $Y_{lm}(\hat{\bm{r}}) \propto e^{im\varphi}$ and integrating over $\varphi$, 
we find that $\Theta_{lm} (l_{\alpha}m_{\alpha},l_{\beta}m_{\beta}) $ is nonzero 
when $m$, $m_{\alpha}$, and $m_{\beta}$ satisfy the relation~\cite{KST69} 
\begin{align}
-m = m_{\alpha} - m_{\beta}, 
\label{eicsmp-eq12}
\end{align}
whereby we can define the integral 
\begin{align}
c^{l} (l_{\alpha}m_{\alpha},l_{\beta}m_{\beta}) 
\equiv  \int d\Omega Y^{*}_{l_{\alpha}m_{\alpha}}(\hat{\bm{r}}) Z_{lm_{\alpha}-m_{\beta}}(\hat{\bm{r}}) Y_{l_{\beta}m_{\beta}}(\hat{\bm{r}}), 
\label{eicsmp-eq13}
\end{align}
which gives 
\begin{align}
\Theta_{lm} (l_{\alpha}m_{\alpha},l_{\beta}m_{\beta})  
 = (-1)^{m} c^{l} (l_{\alpha}m_{\alpha},l_{\beta}m_{\beta}) \delta_{m,m_{\beta}-m_{\alpha}}. 
\label{eicsmp-eq14}
\end{align}
The calculated results for $c^{l} (l_{\alpha}m_{\alpha},l_{\beta}m_{\beta})$ are summarized by Kamimura {\it et al}.~\cite{KST69}, 
where we find that $c^{l} (l_{\alpha}m_{\alpha},l_{\beta}m_{\beta}) $ is nonzero 
when $l$, $l_{\alpha}$, and $l_{\beta}$ satisfy the relations~\cite{KST69} 
\begin{align}
l + l_{\alpha} + l_{\beta} = {\rm even}, \;\;\; |l_{\alpha} - l_{\beta}| \le l \le  l_{\alpha} + l_{\beta}. 
\label{eicsmp-eq15}
\end{align}

Because $Y_{lm}(\hat{\bm{r}})$ and $Z_{lm}(\hat{\bm{r}})$ are complex for $|m|>0$ and real for $m=0$, 
it is convenient to introduce the real spherical harmonics defined as 
\begin{align}
&Y^{(c)}_{lm}(\hat{\bm{r}}) = \frac{1}{ \sqrt{2}}\left[Y_{l-m}(\hat{\bm{r}})+(-1)^{m}Y_{lm}(\hat{\bm{r}}) \right] ,
\label{eicsmp-eq16} \\
&Y^{(s)}_{lm}(\hat{\bm{r}}) = \frac{i}{ \sqrt{2}}\left[Y_{l-m}(\hat{\bm{r}})-(-1)^{m}Y_{lm}(\hat{\bm{r}}) \right] 
\label{eicsmp-eq17}
\end{align} 
for $|m|>0$, which is sometimes called the tesseral harmonics.  
Similarly, we define $Z^{(c)}_{lm}(\hat{\bm{r}})$ and $Z^{(s)}_{lm}(\hat{\bm{r}})$, and 
\begin{align}
&\Theta^{(c)}_{lm} = \frac{1}{\sqrt{2}} \left[ \Theta_{l-m}+(-1)^{m}\Theta_{lm} \right], \\
&\Theta^{(s)}_{lm} = \frac{1}{\sqrt{2}i} \left[ \Theta_{l-m}-(-1)^{m}\Theta_{lm} \right] 
\end{align}
for $|m|>0$.  
The correspondence between the tesseral representation and Cartesian coordinate representation is summarized 
in Table~\ref{eicsmp-table1}, where we find, e.g., that $Y^{(c)}_{11}(\hat{\bm{r}})$ corresponds to the $p_x$ orbital 
and $Y^{(s)}_{22}(\hat{\bm{r}})$ corresponds to the $d_{xy}$ orbital.

\begin{figure}[!t]
\begin{center}
\includegraphics[width=\columnwidth]{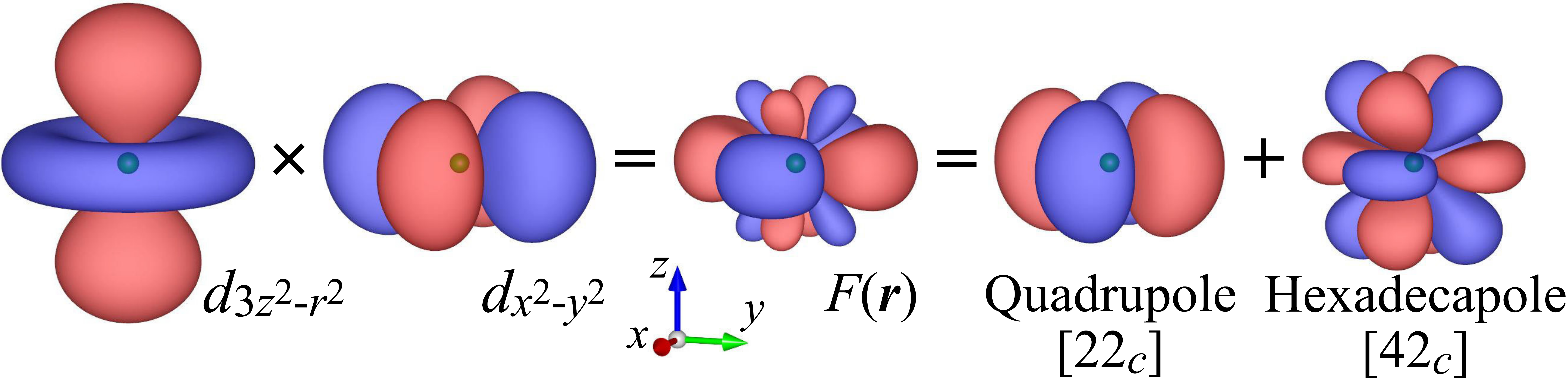}
\caption{(Color online)
Schematic illustration of the multipole expansion of 
$F_{\alpha\beta}(\bm{r})=\phi_{\alpha} (\bm{r})\phi_{\beta} (\bm{r}) $ 
with $\alpha=d_{3z^2-r^2}$ and $\beta=d_{x^2-y^2}$.  
}\label{eicsmp-fig1}
\end{center}
\end{figure}

\begin{table*}[!t]
\begin{center}
{\renewcommand\arraystretch{1.225}
\begin{tabular}{|ll|ll|c|c|c|c|c|}
\hline
$\alpha$ & [$l_{\alpha}m_{\alpha}$] & $\beta$ & [$l_{\beta}m_{\beta}$] & $l=0$ & $l=1$ &$l=2$ & $l=3$ & $l=4$
\\ \hline
$s$ & [00] & $s$   & [00]           & [00] & -- & -- & -- & --
\\ \hline
$s$ & [00] & $p_z$ & [10]      & -- & [10]      & -- & -- & -- \\
    &      & $p_x$ & [11$_c$]  & -- & [11$_c$]  & -- & -- & -- \\
    &      & $p_y$ & [11$_s$]  & -- & [11$_s$]  & -- & -- & -- 
\\ \hline
$s$ & [00] & $d_{3z^2-r^2}$ & [20]     & -- & -- & [20]      & -- & -- \\
    &      & $d_{zx}$       & [21$_c$] & -- & -- & [21$_c$]  & -- & -- \\
    &      & $d_{yz}$       & [21$_s$] & -- & -- & [21$_s$]  & -- & -- \\
    &      & $d_{x^2-y^2}$  & [22$_c$] & -- & -- & [22$_c$]  & -- & --\\
    &      & $d_{xy}$       & [22$_s$] & -- & -- & [22$_s$]  & -- & --    
\\ \hline
$p_z$ & [10] & $p_z$ & [10]          & [00] & -- & [20]           & --  & -- \\
      &      & $p_x$ & [11$_c$]      & --   & -- & [21$_c$]       & --  & -- \\
      &      & $p_y$ & [11$_s$]      & --   & -- & [21$_s$]       & --  & -- \\
$p_x$ & [11$_c$] &  $p_x$ & [11$_c$] & [00] & -- & [20], [22$_c$] & --  & -- \\
      &      &$p_y$ & [11$_s$]       & --   & -- & [22$_s$]       & --  & -- \\
$p_y$ & [11$_s$] &  $p_y$ & [11$_s$] & [00] & -- & [20], [22$_c$] & -- & --  
\\ \hline  
$p_z$ & [10]     & $d_{3z^2-r^2}$ & [20]      & -- & [10]     & -- & [30]     & -- \\
      &          & $d_{zx}$       & [21$_c$]  & -- & [11$_c$] & -- & [31$_c$] & -- \\
      &          & $d_{yz}$       & [21$_s$]  & -- & [11$_s$] & -- & [31$_s$] & -- \\
      &          & $d_{x^2-y^2}$  & [22$_c$]  & -- & --       & -- & [32$_c$] & -- \\
      &          & $d_{xy}$       & [22$_s$]  & -- & --       & -- & [32$_s$] & -- \\  
$p_x$ & [11$_c$] & $d_{3z^2-r^2}$ & [20]      & -- & [11$_c$] & -- & [31$_c$] & -- \\
      &          & $d_{zx}$       & [21$_c$]  & -- & [10]     & -- & [30], [32$_c$] & -- \\
      &          & $d_{yz}$       & [21$_s$]  & -- & --       & -- & [32$_s$] & -- \\
      &          & $d_{x^2-y^2}$  & [22$_c$]  & -- & [11$_c$] & -- & [31$_c$], [33$_c$] & -- \\
      &          & $d_{xy}$       & [22$_s$]  & -- & [11$_s$] & -- & [31$_s$], [33$_s$] & -- \\
$p_y$ & [11$_s$] & $d_{3z^2-r^2}$ & [20]      & -- & [11$_s$] & -- & [31$_s$] & -- \\
      &          & $d_{zx}$       & [21$_c$]  & -- & --       & -- & [32$_s$] & -- \\
      &          & $d_{yz}$       & [21$_s$]  & -- & [10]     & -- & [30], [32$_c$] & -- \\
      &          & $d_{x^2-y^2}$  & [22$_c$]  & -- & [11$_s$] & -- & [31$_s$], [33$_s$] & -- \\
      &          & $d_{xy}$       & [22$_s$]  & -- & [11$_c$] & -- & [31$_c$], [33$_c$] & --  
\\ \hline
$d_{3z^2-r^2}$ & [20]     & $d_{3z^2-r^2}$ & [20]     & [00] & -- & [20]           & -- & [40]  \\
               &          & $d_{zx}$       & [21$_c$] & --   & -- & [21$_c$]       & -- & [41$_c$]  \\
               &          & $d_{yz}$       & [21$_s$] & --   & -- & [21$_s$]       & -- & [41$_s$]  \\
               &          & $d_{x^2-y^2}$  & [22$_c$] & --   & -- & [22$_c$]       & -- & [42$_c$]  \\
               &          & $d_{xy}$       & [22$_s$] & --   & -- & [22$_s$]       & -- & [42$_s$]  \\  
$d_{zx}$       & [21$_c$] & $d_{zx}$       & [21$_c$] & [00] & -- & [20],[22$_c$]  & -- & [40], [42$_c$]  \\
               &          & $d_{yz}$       & [21$_s$] & --   & -- & [22$_s$]       & -- & [42$_s$]  \\
               &          & $d_{x^2-y^2}$  & [22$_c$] & --   & -- & [21$_c$]       & -- & [41$_c$], [43$_c$]  \\
               &          & $d_{xy}$       & [22$_s$] & --   & -- & [21$_s$]       & -- & [41$_s$], [43$_s$]  \\ 
$d_{yz}$       & [21$_s$] & $d_{yz}$       & [21$_s$] & [00] & -- & [20], [22$_c$] & -- & [40], [42$_c$]  \\
               &          & $d_{x^2-y^2}$  & [22$_c$] & --   & -- & [21$_s$]       & -- & [41$_s$], [43$_s$]  \\
               &          & $d_{xy}$       & [22$_s$] & --   & -- & [21$_c$]       & -- & [41$_c$], [43$_c$] \\ 
$d_{x^2-y^2}$  & [22$_c$] & $d_{x^2-y^2}$  & [22$_c$] & [00] & -- & [20]           & -- & [40], [44$_c$]   \\
               &          & $d_{xy}$       & [22$_s$] & --   & -- & --             & -- & [44$_s$]  \\ 
$d_{xy}$       & [22$_s$] & $d_{xy}$       & [22$_s$] & [00] & -- & [20]           & -- & [40], [44$_c$]                
\\ \hline    
\end{tabular}}
\caption{Correspondence between the orbitals $\alpha$ and $\beta$ and the nonvanishing multipole moments, 
where $[lm_c]$ and $[lm_s]$ indicate  $Y^{(c)}_{lm}(\hat{\bm{r}})$ [or $Z^{(c)}_{lm}(\hat{\bm{r}})$] and 
$Y^{(s)}_{lm}(\hat{\bm{r}})$ [or $Z^{(s)}_{lm}(\hat{\bm{r}})$], respectively. }
\label{eicsmp-table2}
\end{center}
\end{table*}

Next, let us evaluate $\Theta_{lm}$ in the tesseral representation.  
The results for the orbitals from $s$ ($l=0$) to $d$ ($l=2$) are summarized in Table~\ref{eicsmp-table2}.  
Here, we describe an example, where $\alpha$ is the $d_{3z^2-r^2}$ ($[l_{\alpha}m_{\alpha}]=[20]$) orbital 
and $\beta$ is the $d_{x^2-y^2}$ ($[l_{\beta}m_{\beta}]=[22_c]$) orbital.  
Using $l_{\alpha}=l_{\beta}=2$ and the relation in Eq.~(\ref{eicsmp-eq15}), 
we find that the ranks of possible multipoles are $l=0$, $2$, and $4$.  
Using the tesseral harmonics, 
we find 
\begin{align}
\Theta_{lm}(d_{3z^2-r^2},d_{x^2-y^2}) 
&= (-1)^m \int d\Omega Y_{20}  Z_{l-m} Y^{(c)}_{22} 
\notag \\
&= \frac{1}{\sqrt{2}} c^l (20,2\pm2)\delta_{m,\pm2} , 
\label{eicsmp-eq18}
\end{align}
which indicates that $\Theta_{lm}\ne0$ at $m=\pm 2$.  
Therefore, $\Theta_{lm}$ with $l=2$ and $4$ and $m=\pm 2$ remain and are given by 
\begin{align}
\Theta_{2\pm2}(d_{3z^2-r^2},d_{x^2-y^2}) 
&= \frac{1}{\sqrt{2}} c^2 (20,2\pm2) 
= -\frac{\sqrt{2}}{7} ,
\label{eicsmp-eq19}\\
\Theta_{4\pm2}(d_{3z^2-r^2},d_{x^2-y^2}) 
&= \frac{1}{\sqrt{2}} c^4 (20,2\pm2) 
= \frac{\sqrt{15}}{21\sqrt{2}}, 
\label{eicsmp-eq20}
\end{align}
where we use $c^2 (20,2\pm2)=-2/7$ and $c^4 (20,2\pm2)=\sqrt{15}/21$~\cite{KST69}.  
Applying the tesseral representation to $\Theta_{l\pm2}$, we find 
\begin{align}
&\Theta^{(s)}_{l2}(d_{3z^2-r^2},d_{x^2-y^2}) \propto \left[ \Theta_{l-2}-\Theta_{l2} \right]=0, 
\end{align}
\begin{align}
&\Theta^{(c)}_{22}(d_{3z^2-r^2},d_{x^2-y^2}) = \frac{1}{ \sqrt{2}}\left[\Theta_{2-2}+\Theta_{22} \right]  = -\frac{2}{7} ,
\label{eicsmp-eq21}\\
&\Theta^{(c)}_{42}(d_{3z^2-r^2},d_{x^2-y^2}) = \frac{1}{ \sqrt{2}}\left[\Theta_{4-2}+\Theta_{42} \right]  =\frac{\sqrt{15}}{21} .
\label{eicsmp-eq22}
\end{align}
Thus, the quadrupole moment $Q^{(c)}_{22}=Q_{x^2-y^2}$ and hexadecapole moment $Q^{(c)}_{42}=Q_{(x^2-y^2)(7z^2-r^2)}$ remain finite 
when the $d_{3z^2-r^2}$ and $d_{x^2-y^2}$ orbitals are hybridized spontaneously. 

Finally, let us consider the multipole expansion of the product of the wave functions 
\begin{align}
F_{\alpha\beta}(\bm{r}) =  \phi^{*}_{\alpha} (\bm{r})\phi_{\beta} (\bm{r}) .
\label{eicsmp-eq23}
\end{align}
given in Eq.~(\ref{eicsmp-eq3}).  
Using $\sum_{l,m}(2l+1) Z^{*}_{lm}(\hat{\bm{r}}') Z_{lm}(\hat{\bm{r}}) = 4\pi\delta(\hat{\bm{r}}-\hat{\bm{r}}')$, 
we have
\begin{align}
&Y^{*}_{l_{\alpha}m_{\alpha}}(\hat{\bm{r}}) Y_{l_{\beta}m_{\beta}}(\hat{\bm{r}})  \notag \\
&\;\;\; = \sum_{l,m} \left(\frac{2l+1}{4\pi}\right) \Theta_{lm} (l_{\alpha}m_{\alpha},l_{\beta}m_{\beta})  Z_{lm}(\hat{\bm{r}}). 
\label{eicsmp-eq24}
\end{align}
Therefore, from Eqs.~(\ref{eicsmp-eq2}) and (\ref{eicsmp-eq24}), the multipole expansion of $F_{\alpha\beta}(\bm{r})$ is given by 
\begin{align}
F_{\alpha\beta}(\bm{r}) 
&= R_{n_{\alpha}l_{\alpha}}(r) R_{n_{\beta}l_{\beta}}(r) 
\notag \\
&\times \sum_{l,m} \left(\frac{2l+1}{4\pi}\right) \Theta_{lm} (l_{\alpha}m_{\alpha},l_{\beta}m_{\beta})  Z_{lm}(\hat{\bm{r}}). 
\label{eicsmp-eq25}
\end{align}
Using Eq.~(\ref{eicsmp-eq25}), we finally obtain the electronic density in Eq.~(\ref{eicsmp-eq3}) as 
\begin{align}
&\rho(\bm{r}) 
= \sum_{\alpha,\beta} R_{n_{\alpha}l_{\alpha}}(r) R_{n_{\beta}l_{\beta}}(r) \notag \\
& \times \left[ \sum_{l,m} \left(\frac{2l+1}{4\pi}\right) \Theta_{lm} (l_{\alpha}m_{\alpha},l_{\beta}m_{\beta})  Z_{lm}(\hat{\bm{r}}) \right] \langle c^{\dag}_{\alpha}c_{\beta} \rangle. 
\label{eicsmp-eq26}
\end{align}

In the previous example with $\alpha=d_{3z^2-r^2}$ and $\beta=d_{x^2-y^2}$, where we find $\Theta^{(c)}_{22}\ne 0$ and $\Theta^{(c)}_{42}\ne 0$, 
we can expand $F_{\alpha\beta}(\bm{r}) $ into the quadrupole and hexadecapole as 
\begin{align}
F_{\alpha\beta}(\bm{r}) 
&= \frac{5}{4\pi} R^2_{32}(r) \Theta^{(c)}_{22} (d_{3z^2-r^2},d_{x^2-y^2})  Z^{(c)}_{22}(\hat{\bm{r}})
\notag \\
&+ \frac{9}{4\pi} R^2_{32}(r) \Theta^{(c)}_{42} (d_{3z^2-r^2},d_{x^2-y^2})  Z^{(c)}_{42}(\hat{\bm{r}}), 
\label{eicsmp-eq27}
\end{align}
of which the schematic illustration of the multipole expansion is given in Fig.~\ref{eicsmp-fig1}.  
Thus, the product of the wave functions $F_{\alpha\beta}(\bm{r})$ is given in general by the sum of the multipoles.  

\end{appendix}

\bibliography{Paper}
\end{document}